\documentclass[10pt,aps,prd,nofootinbib,superscriptaddress,floatfix]{revtex4-2}
\usepackage[utf8]{inputenc}
\usepackage{amsmath}
\usepackage{amssymb}
\usepackage{graphicx}
\graphicspath{{Figures/}}
\usepackage{subcaption}
\usepackage{gensymb}
\usepackage[bookmarks=false,
 breaklinks=false,pdfborder={0 0 1},backref=section,colorlinks=false]{hyperref}
\hypersetup{colorlinks,bookmarksopen,bookmarksnumbered,linkcolor=blu,pdfstartview=FitH,urlcolor=rossos,citecolor=rossos}
\usepackage{color}
\usepackage{textcomp}\usepackage{amsfonts}\usepackage{graphics}\usepackage{epstopdf}\usepackage{slashed}\usepackage{multirow}\usepackage{adjustbox}\usepackage{lipsum}\usepackage{rotating}\usepackage{xcolor}\usepackage{wrapfig}\usepackage{epsfig}\usepackage{tikzsymbols}\usepackage{tikz}
\definecolor{rosso}{cmyk}{0,1,1,0.4}
\definecolor{rossos}{cmyk}{0,1,1,0.55}
\definecolor{rossoc}{cmyk}{0,1,1,0.2}
\definecolor{blu}{cmyk}{1,1,0,0.3}
\definecolor{blus}{cmyk}{1,1,0,0.6}
\definecolor{bluc}{cmyk}{1,1,0,0.1}
\definecolor{verde}{cmyk}{0.92,0,0.59,0.25}
\definecolor{verdec}{cmyk}{0.92,0,0.59,0.15}
\definecolor{verdes}{cmyk}{0.92,0,0.59,0.4}
\definecolor{RED}{rgb}{1,0,0}
\definecolor{BLUE}{rgb}{0,0,1}

\definecolor{lime}{HTML}{A6CE39}
\DeclareRobustCommand{\orcidicon}{
	\begin{tikzpicture}
	\draw[lime, fill=lime] (0,0)
	circle [radius=0.2]
	node[white] {{\fontfamily{qag}\selectfont \tiny ID}};
	\draw[white, fill=white] (-0.0625,0.095)
	circle [radius=0.007];
	\end{tikzpicture}
	\hspace{-2mm} }
\foreach \x in {A, ..., Z}{\expandafter\xdef\csname orcid\x\endcsname{\noexpand\href{https://orcid.org/\csname orcidauthor\x\endcsname}
			{\noexpand\orcidicon}} }

\definecolor{lime}{HTML}{A6CE39}
\foreach \x in {A, ..., Z}{\expandafter\xdef\csname orcid\x\endcsname{\noexpand\href{https://orcid.org/\csname orcidauthor\x\endcsname}
			{\noexpand\orcidicon}} }

\newcommand{\arXivold}[2]{\href{http://arxiv.org/pdf/#1}{{\tt #2/#1}}}

\begin{document}
\title{Detectability of Magnetar-Induced Vacuum Birefringence with IXPE and eXTP}
\author{Fayez Abu-Ajamieh\orcidA{}}
\email{fayezabuajamieh@gmail.com}

\affiliation{Formerly: Center for High Energy Physics; Indian Institute of Science;
Bangalore; India}

\begin{abstract}
We analyze finite-field vacuum birefringence from magnetars and its implications for X-ray polarization measurements with IXPE and eXTP. We adopt a realistic spatially-varying magnetic field profile for magnetars and use the one-loop refractive indices, without expanding in $B/B_{\rm c}$, to calculate the differential time delay and accumulated phase between the two photon polarization eigenmodes. Including the extended magnetosphere enhances the time delay by a factor of 1.2--1.7 relative to a constant-field estimate and can yield delays approaching the microsecond scale for the strongest catalogued magnetar fields. We also study the evolution of the polarization state in a varying projected magnetic field using the Stokes-vector precession equation, derive the corresponding adiabaticity condition and polarization-limiting radius, and recover the fixed-eigenaxis expressions used in our catalog analysis as a controlled limit. We calculate the Stokes parameters for catalogued magnetar fields and investigate the effects of instrumental response and energy averaging. The fixed-eigenaxis phase oscillations average away over realistic X-ray energy bands, so the resulting polarization-measurement sensitivity does not by itself establish the detectability of vacuum birefringence. A quantitative prediction for an individual magnetar, however, additionally requires a source-specific treatment of the emission mechanism, viewing geometry, plasma environment, and magnetospheric scattering, which we do not tackle in this study.

\end{abstract}
\maketitle

\section{Introduction}\label{sec1}
It has long been known that Quantum Electrodynamics (QED) behaves differently in the presence of a strong electric or magnetic background field compared to weak background fields. The first attempt at describing this behavior was given by Euler-Heisenberg~\cite{Heisenberg:1936nmg} and Weisskopf~\cite{Weisskopf:1936}, and later by Schwinger using the proper-time formalism~\cite{Schwinger:1951nm}. The Euler-Heisenberg Lagrangian describes the one-loop effective Lagrangian density of QED in the presence of a constant background field. It encodes nonlinear vacuum polarization effects and, in the presence of electric fields, an imaginary part corresponding to Schwinger pair production:
\begin{equation}\label{eq:EH_Lag}
\mathcal{L}_{\text{EH}}(\mathbf{E},\mathbf{B}) = -\frac{1}{8\pi^{2}}\int_{0}^{\infty}\frac{ds}{s^{3}}e^{-m^{2}s}\Bigg[(es)^{2}\mathcal{G} \frac{\text{Re}\cosh({es\sqrt{2(\mathcal{F}+i\mathcal{G})}})}{\text{Im}\cosh{(es\sqrt{2(\mathcal{F}+i\mathcal{G})}})} -\frac{2}{3}(es)^{2}\mathcal{F}-1\Bigg],
\end{equation}
where $m$ is the mass of the electron and
\begin{align}
\mathcal{F} & = \frac{1}{4}F_{\mu\nu}F^{\mu\nu} = \frac{1}{2}(\textbf{B}^{2}-\textbf{E}^{2}),\label{eq:F}\\
\mathcal{G} & =  \frac{1}{4}F_{\mu\nu}\tilde{F}^{\mu\nu} = (\textbf{E}\cdot\textbf{B}), \hspace{1cm} \tilde{F}^{\mu\nu} = \frac{1}{2} \epsilon_{\mu\nu\rho\sigma}F^{\rho\sigma}. \label{eq:G}
\end{align}

The presence of a strong background field can give rise to phenomena like birefringence~\cite{Heisenberg:1936nmg, Weisskopf:1936}, Delbr{\"u}ck scattering~\cite{Meitner:1933kww, Delbruck:1933pla}, where a photon scatters off a strong Coulomb field, Schwinger pair production~\cite{Sauter:1931zz, Schwinger:1951nm}, Light-By-Light (LBL) scattering, and photon splitting~\cite{Adler:1971wn}. The Delbr{\"u}ck scattering has been observed~\cite{Moreh:1973gma, Rullhusen:1983zz, Jarlskog:1973aui} and photon splitting has also been observed~\cite{Jarlskog:1973aui}. In addition, heavy-ion collisions were proposed as a probe for LBL scattering~\cite{dEnterria:2013zqi}, with evidence for it presented by ATLAS~\cite{ATLAS:2017fur} and CMS~\cite{CMS:2018erd}, and with proposals put forward for its detection in future $e^{+}e^{-}$ colliders~\cite{Ellis:2022uxv} and muon colliders~\cite{Yang:2020rjt, Amarkhail:2023xsc, Spor:2024nsx}. On the other hand, Schwinger pair production remains elusive. This can be easily understood by inspecting the rate of Schwinger pair production in the presence of a constant electric field, which is given by (in units where $c = \hbar = 1$)
\begin{equation}\label{eq:Schwinger_rate}
\Gamma(E\rightarrow e^{+}e^{-}) = \frac{\alpha E^{2}}{2\pi^{2}}\mathrm{Li_{2}}\Big( e^{-\frac{\pi m^{2}}{eE}}\Big),
\end{equation}
where the critical field $E_{\text{c}} = m^{2}c^{3}/e\hbar \approx 1.32 \times 10^{18}~\text{V/m}$. Similarly, the critical magnetic field is given by $B_{\text{c}} = m^{2}c^{2}/e\hbar \approx 4.41 \times 10^{13}~\text{G}$. Thus, we see that the rate of Schwinger pair production for $E \ll E_{\text{c}}$ is exponentially suppressed. High-intensity lasers can be used to create strong $E$ fields that probe the dynamically-assisted Schwinger pair production. However, the highest laser intensity of $1.1\times 10^{23}~\text{W}/\text{cm}^{2}$, which was achieved by the Center for Relativistic Laser Science (CoReLS) laser~\cite{Yoon:2021ony}, only leads to an electric field three orders of magnitude smaller than $E_{\text{c}}$. Lasers with a higher intensity are proposed~\cite{Danson:2019qlu}, but they only increase the achievable electric field by 1-2 orders of magnitude.

Vacuum birefringence, which refers to the different phase velocities/refractive indices for different polarization eigenmodes, is another key phenomenon that arises in a strong background field, with experiments designed for its detection. The Polarizzazione del Vuoto con LASer (PVLAS) experiment~\cite{DellaValle:2015xxa} was designed to utilize a permanent magnet to create the magnetic background, and a high-intensity laser as probe, in order to probe vacuum birefringence. However, the PVLAS experiment was unable to find evidence of vacuum birefringence to date.

The difficulty in creating ultra-high background fields in terrestrial experiments motivates using neutron stars as strong-field laboratories. Magnetar surface fields can reach $\sim 10$--$100B_{\text{c}}$. The theory of photon propagation in this setting is extensive. Arbitrary-field QED indices and dichroism were analyzed in Ref.~\cite{HeylHernquist:1997}; polarization-dependent magnetic lensing in a dipole field in Ref.~\cite{ShavivHeylLithwick:1999}; and the adiabatic evolution and freeze-out of the photon polarization state in Refs.~\cite{HeylShaviv:1999,HeylShaviv:2002}. The plasma--vacuum resonance and mode conversion in a magnetized atmosphere were developed in Refs.~\cite{LaiHo:2003,HoLai:2003}, followed by full modal and Stokes radiative-transfer calculations~\cite{vanAdelsbergLai:2006}. Propagation through quasi-tangential regions, surface integration, general-relativistic ray bending and magnetospheric scattering were subsequently studied in Refs.~\cite{WangLai:2009,FernandezDavis:2011,Taverna:2014,Kelly:2024}.

The Imaging X-ray Polarimetry Explorer (IXPE) is a NASA-led mission, developed in collaboration with the Italian Space Agency, designed to measure the linear polarization of astrophysical X-ray sources in the 2–8 keV energy range. Launched in December 2021, IXPE has provided polarization measurements of several magnetars, offering a new observational window into strong-field QED effects. The enhanced X-ray Timing and Polarimetry mission (eXTP) is a next-generation X-ray observatory under development, with a planned larger collecting area and complementary timing, spectroscopic, and polarimetric capabilities. Its Polarimetry Focusing Array is designed to improve the sensitivity of X-ray polarization measurements, potentially enabling more precise investigations of magnetar magnetospheres.

IXPE has now detected significant and diverse X-ray polarization from several magnetars. The first result for 4U~0142+61 showed a strongly energy-dependent signal~\cite{Taverna:2022jgl}; 1RXS~J170849.0$-$400910 exhibited a larger polarization fraction and nearly constant angle~\cite{Zane:2023}; and later observations further demonstrated the importance of emission components and viewing geometry~\cite{TavernaTurolla:2024}. Recent analyses of 1E~1547.0$-$5408 illustrate the remaining model dependence: an atmosphere-plus-geometry analysis found compelling evidence for magnetospheric vacuum birefringence~\cite{Stewart:2025}, whereas an independent analysis emphasized that a small emitting spot and inclined geometry can produce high polarization without a unique vacuum-birefringence inference~\cite{Taverna:2026}. These measurements strongly motivate QED propagation calculations, but do not directly measure the refractive-index difference without an emission and geometry model.

In this paper we focus on a narrower propagation question. We adopt a spatially varying dipole-magnitude profile and use the one-loop expressions without expanding in $B/B_{\text{c}}$. We calculate the differential delay and phase for a distant background source whose light follows a grazing trajectory through the magnetosphere, comparing the result with weak-field and constant-field estimates. We then formulate the general evolution of the Stokes vector when the local polarization eigenaxes rotate along the trajectory. The catalog calculations are retained as controlled, fixed-eigenaxis benchmarks rather than source-specific models of thermal emission. They illustrate finite-field phase accumulation and detector band averaging, while exposing the additional ingredients required for a quantitative comparison with IXPE or eXTP. The latter mission is currently planned for launch in 2030~\cite{eXTP:2025}.

Exact finite-field corrections have also been applied to the polarization-limiting radius and the atmospheric plasma--vacuum resonance~\cite{Abbassi:2026}, while nonlinear-electrodynamic corrections to ray tracing and travel time have been considered in Ref.~\cite{Porto:2026}. Reference~\cite{Abbassi:2026} finds that finite-field corrections are most important close to the stellar surface and at the vacuum resonance, whereas the polarization-limiting radius lies far enough out that the weak-field form is already accurate. Our propagation calculation is complementary: it retains the full near-field phase and delay along an external grazing ray, but it does not replace atmosphere or magnetospheric radiative transfer.

This paper is organized as follows. Section~\ref{sec2} reviews vacuum birefringence and the exact and weak-field indices. Section~\ref{sec3} calculates the group delay and phase along the idealized grazing trajectory. Section~\ref{sec4} develops polarization-state evolution, identifies the fixed-eigenaxis limit and presents response-weighted benchmarks. We conclude in Section~\ref{sec5}.

\section{Vacuum Birefringence}\label{sec2}
Vacuum birefringence refers to the phenomenon where the index of refraction in the presence of a magnetic background field is different for each polarization eigenmode. More specifically, in the presence of a strong magnetic field, the index of refraction in the direction parallel to the ($\mathbf{B}$,$\mathbf{k}$) plane, will not be equal to the index of refraction in the direction perpendicular to it, i.e., $n_{\parallel} \neq n_{\perp}$, which implies that there will be a difference in the phase velocity of the light components with parallel and perpendicular polarizations. A comprehensive treatment of birefringence in the presence of a \textit{constant} background field was provided by Adler~\cite{Adler:1971wn}, who calculated the one-loop correction to the photon propagator with any number of background photon insertions in the loop,
\begin{equation}\label{eq:Adler1}
n_{\parallel,\perp} = 1-\frac{1}{2}\sin^{2}{\theta}A_{\parallel,\perp}(\omega,B),
\end{equation}
where $\omega$ is the energy of the incident photon and $\theta\equiv\theta_{kB}$ is the local angle between $\mathbf B$ and the propagation direction. The two linear eigenvectors are defined locally: one lies in the $(\mathbf B,\mathbf k)$ plane and the other is perpendicular to it. Consequently, a spatially varying vector field changes both the eigenvalues and the orientation of the eigenvectors. A scalar phase integral alone is sufficient only when the projected field direction is fixed. The corresponding polarization transport is developed in Section~\ref{sec4}. The functions appearing in Eq.~(\ref{eq:Adler1}) are
\begin{align}
A_{\parallel,\perp}(\omega,B) & = \frac{\alpha}{2\pi}\int_{0}^{\infty}\frac{ds}{s^{2}}\exp{(-m^{2}s)}\int_{0}^{s}dt \exp[\omega^{2}R(s,t)]J_{\parallel,\perp}(s,v), \label{eq:Adler2}\\
J_{\parallel}(s,v) & =\frac{-eBs\cosh{(eBsv)}}{\sinh{(eBs)}}+ \frac{eBsv\sinh{(eBsv)}\coth{(eBs)}}{\sinh{(eBs)}} - \frac{2eBs[\cosh{(eBsv)}-\cosh{(eBs)}]}{\sinh^{3}{(eBs)}},\label{eq:Adler3}\\
J_{\perp}(s,v) & = \frac{eBs\cosh{(eBsv)}}{\sinh{(eBs)}}- eBs\coth{(eBs)}\Big[1-v^{2}+v + \frac{\sinh{(eBsv)}}{\sinh{(eBs)}}\Big],\label{eq:Adler4}\\
R(s,t) & = \frac{1}{2}\Big[2t\Big(1-\frac{t}{s}\Big) + \frac{\cosh{[eBsv]}-\cosh{(eBs)}}{eB\sinh{(eBs)}}\Big],
\end{align}
where $v = 2t/s -1$. In the limit of $\omega \rightarrow 0$, an analytic solution exists for an arbitrary constant magnetic field~\cite{Dittrich:1998fy, Kim:2021kif}, which in the notation of the latter is given by (for $\theta = \frac{\pi}{2}$)
\begin{equation}\label{eq:exact_sol1}
n_{\parallel} = \sqrt{\frac{1-X+Y}{1-X}}, \hspace{5mm} n_{\perp} = \sqrt{\frac{1-X}{1-X-Z}},
\end{equation}
where
\begin{align}
X & =-\frac{\alpha}{6\pi}[1 - 6\ln{(2\pi)}\overline{a} + 6\overline{a}^{2} + 2\ln{\overline{a}}-6\overline{a}\ln\overline{a} + 12\overline{a}\ln\Gamma(\overline{a})-24\zeta'(-1,\overline{a})],\label{eq:exact_sol2}\\
Y & =-\frac{\alpha}{6\pi \overline{a}}[1+ \overline{a}-6\ln(2\pi)\overline{a}^{2}+6\overline{a}^{3}-6\overline{a}^{2}\ln\overline{a}+12\overline{a}^{2}\ln\Gamma(\overline{a})+2\overline{a}\psi(\overline{a})-24\overline{a}\zeta'(-1,\overline{a})],\label{eq:exact_sol3}\\
Z & = \frac{\alpha}{3\pi}[1+3(1+\ln(2\pi))\overline{a}-6\overline{a}^{2}-3\overline{a}\ln{\overline{a}}-6\overline{a}\ln{\Gamma(\overline{a})} + 6\overline{a}^{2}\psi(\overline{a})],\label{eq:exact_sol4}
\end{align}
and $\zeta'(-1,\overline{a}) = d\zeta(s,\overline{a})/ds|_{s=-1}$ and $\zeta(s,\overline{a})$ is the Hurwitz zeta function, $\Gamma(\overline{a})$ and $\psi(\overline{a})$ are the gamma and digamma functions, respectively, $\overline{a} = m^{2}/(2ea)$ which for a pure magnetic field simplifies to $\overline{a} =  B_{\text{c}}/(2B)$; and the Lorentz-invariant quantity $a$ is given by
\begin{equation}
a \equiv \sqrt{\sqrt{\mathcal{F}^{2}+ \mathcal{G}^{2}}+\mathcal{F}}.
\end{equation}
The exact refractive indices can be expanded in the weak-field limit, \(B\ll B_c\). Retaining the leading-order (LO) and next-to-leading-order (NLO) contributions, one obtains
\begin{align}
n_{\parallel} & = 1 + \frac{14\alpha^{2}}{45m^{4}}\Big(1+\frac{1315\alpha}{252\pi}\Big)B^{2},\label{eq:n_approx_1}\\
n_{\perp} & = 1 + \frac{8\alpha^{2}}{45m^{4}}\Big(1 +\frac{40\alpha}{9\pi} \Big)B^{2},\label{eq:n_approx_2}
\end{align}
and one can clearly see how a magnetic field induces birefringence. In particular, we find that 
\begin{equation}\label{eq:Delta_n}
\Delta n = n_{\parallel} - n_{\perp} \approx \frac{2}{15}\Big( \frac{\alpha B}{m^{2}} \Big)^{2}\Big(1+\frac{25\alpha}{4\pi}\Big).
\end{equation}

\begin{figure}[t!]
\includegraphics[width=0.6\textwidth]{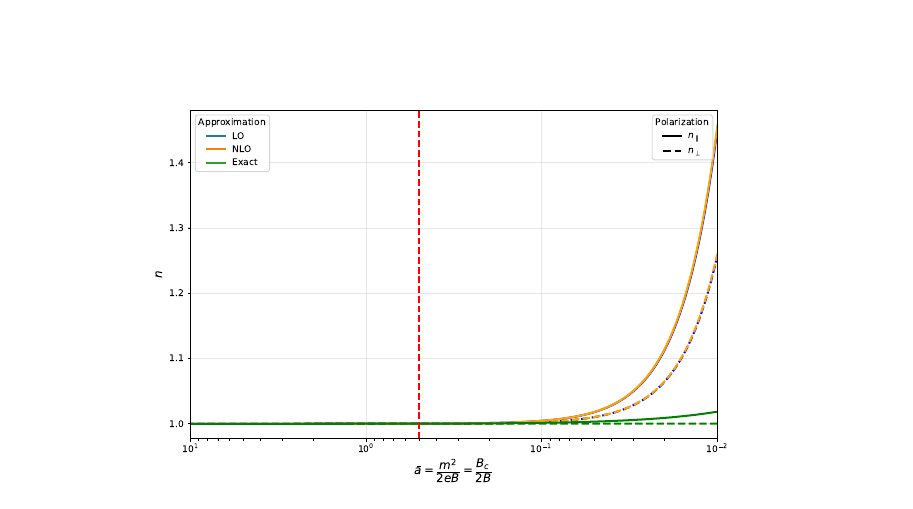}
 \caption{{\small{$n_{\parallel}$ (solid) and $n_{\perp}$ (dashed) from the LO (blue), NLO (orange) and exact solution (green) against $\overline{a} = B_{c}/2B$. The vertical line corresponds to the critical field.}}}
\label{fig:n_comparison} 
\end{figure}

We plot $n_{\parallel}$ and $n_{\perp}$ from the exact solution, in addition to the LO and NLO approximations in Figure~\ref{fig:n_comparison}. The plot shows several important features. First, significant departure from unity only takes place for $B \gtrsim B_{\text{c}}$, whereas it remains negligible for smaller field strength. Second, $n_{\parallel}$ grows faster than $n_{\perp}$, which although evident from the LO and NLO expansions in Eqs.~(\ref{eq:n_approx_1}) and~(\ref{eq:n_approx_2}), is not so obvious from the exact solution. Third, there is little difference between the LO and NLO expansions. The NLO formula includes a higher-loop correction, whereas the exact finite-field result is one-loop. Finally and most importantly, the LO and NLO approximations tend to overestimate the level of reduction in the speed of light for $B \gtrsim B_{\text{c}}$. For instance, the LO expansion suggests that the phase velocity is reduced to less than half its vacuum value for $B \sim 100B_{\text{c}}$, however, the exact solution suggests a reduction of only a few percent. This indicates that for strong magnetic fields like in magnetars, the expansion breaks down and the full solution must be taken into account

\section{Vacuum Birefringence from Magnetars}\label{sec3}
We first consider a distant background source whose ray passes through the magnetosphere of a foreground magnetar. This is distinct from the surface-emission geometry relevant to the observed persistent X-rays of the magnetar itself. A propagation observable for the background-source configuration is $\Delta t$, the relative arrival-time delay of the two eigenmodes. Previous studies found delays in the approximate range $10^{-8}$--$10^{-7}$ s for representative pulsar and magnetar parameters~\cite{Denisov:2005si,Denisov:2014oka,Abishev:2014ceb,Denisov:2016pfu,Abishev:2018ahd}. Strictly, a wave-packet delay is controlled by the group index,
\begin{equation}\label{eq:delta_t}
\Delta t = \frac{L}{c}\Delta n_g,\qquad
\Delta n_g=\Delta n+\omega\frac{\partial\Delta n}{\partial\omega}.
\end{equation} 
Here $L$ is the propagation length. In the low-energy range considered below, $\omega\ll m_e$, the indices are nondispersive to very good accuracy and $\Delta n_g\simeq\Delta n$. We use the field strengths in the McGill Online Magnetar Catalog~\cite{McGill:2026} as a parameter set. Table~\ref{tab1} first takes $L=R_M=10$ km and compares weak-field and finite-field expressions. The weak-field expansion is reliable for $B/B_{\text c}\lesssim\mathcal O(1)$ but increasingly overestimates the delay at supercritical fields. The finite-field values are compatible in scale with Refs.~\cite{Denisov:2005si,Denisov:2014oka,Abishev:2014ceb,Denisov:2016pfu,Abishev:2018ahd}. A constant field truncated at $R_M$, however, omits the extended magnetosphere; the following calculation quantifies that geometric correction.
\begin{table}[t!]
\centering
\small
\begin{tabular}{lcccccc}
\hline
\textbf{Magnetar Name} & $\bf{B\,[\times 10^{15}\,\mathrm{G}]}$ & $\bf{B/B_c}$ & $\bf{\Delta t_{\rm LO}\,[\mathrm{s}]}$ & $\bf{\Delta t_{\rm NLO}\,[\mathrm{s}]}$ & $\bf{\Delta t_{\rm Exact}\,[\mathrm{s}]}$ & $\bf{\Delta t_{\rm Model}\,[\mathrm{s}]}$ \\
\hline

SGR 1806$-$20 & 2.00 & 45.5 & $5.3\times10^{-6}$ & $5.4\times10^{-6}$ & $5.5\times10^{-7}$ & $9.6\times10^{-7}$ \\
SGR 1900+14 & 0.70 & 15.9 & $6.5\times10^{-7}$ & $6.6\times10^{-7}$ & $1.7\times10^{-7}$ & $2.8\times10^{-7}$ \\
1E 1841$-$045 & 0.70 & 15.9 & $6.5\times10^{-7}$ & $6.6\times10^{-7}$ & $1.7\times10^{-7}$ & $2.8\times10^{-7}$ \\
SGR 0526$-$66 & 0.56 & 12.7 & $4.2\times10^{-7}$ & $4.2\times10^{-7}$ & $1.3\times10^{-7}$ & $2.1\times10^{-7}$ \\
CXOU J171405.7$-$381031 & 0.50 & 11.4 & $3.3\times10^{-7}$ & $3.4\times10^{-7}$ & $1.2\times10^{-7}$ & $1.8\times10^{-7}$ \\
1RXS J170849.0$-$400910 & 0.47 & 10.7 & $2.9\times10^{-7}$ & $3.0\times10^{-7}$ & $1.1\times10^{-7}$ & $1.7\times10^{-7}$ \\
CXOU J010043.1$-$721134 & 0.39 & 8.9 & $2.0\times10^{-7}$ & $2.1\times10^{-7}$ & $8.7\times10^{-8}$ & $1.3\times10^{-7}$ \\
1E 1048.1$-$5937 & 0.39 & 8.9 & $2.0\times10^{-7}$ & $2.1\times10^{-7}$ & $8.7\times10^{-8}$ & $1.3\times10^{-7}$ \\
Swift J1818.0$-$1607 & 0.35 & 8.0 & $1.6\times10^{-7}$ & $1.7\times10^{-7}$ & $7.6\times10^{-8}$ & $1.1\times10^{-7}$ \\
1E 1547.0$-$5408 & 0.32 & 7.3 & $1.4\times10^{-7}$ & $1.4\times10^{-7}$ & $6.8\times10^{-8}$ & $9.9\times10^{-8}$ \\
PSR J1622$-$4950 & 0.27 & 6.1 & $9.7\times10^{-8}$ & $9.9\times10^{-8}$ & $5.4\times10^{-8}$ & $7.8\times10^{-8}$ \\
SGR J1745$-$2900 & 0.23 & 5.2 & $7.1\times10^{-8}$ & $7.2\times10^{-8}$ & $4.4\times10^{-8}$ & $6.1\times10^{-8}$ \\
SGR 1627$-$41 & 0.22 & 5.0 & $6.5\times10^{-8}$ & $6.6\times10^{-8}$ & $4.1\times10^{-8}$ & $5.7\times10^{-8}$ \\
SGR 1935+2154 & 0.22 & 5.0 & $6.5\times10^{-8}$ & $6.6\times10^{-8}$ & $4.1\times10^{-8}$ & $5.7\times10^{-8}$ \\
XTE J1810$-$197 & 0.21 & 4.8 & $5.9\times10^{-8}$ & $6.0\times10^{-8}$ & $3.9\times10^{-8}$ & $5.4\times10^{-8}$ \\
SGR 0501+4516 & 0.19 & 4.3 & $4.8\times10^{-8}$ & $4.9\times10^{-8}$ & $3.4\times10^{-8}$ & $4.6\times10^{-8}$ \\
SGR 1833$-$0832 & 0.16 & 3.6 & $3.4\times10^{-8}$ & $3.5\times10^{-8}$ & $2.6\times10^{-8}$ & $3.5\times10^{-8}$ \\
Swift J1834.9$-$0846 & 0.14 & 3.2 & $2.6\times10^{-8}$ & $2.7\times10^{-8}$ & $2.2\times10^{-8}$ & $2.8\times10^{-8}$ \\
4U 0142+61 & 0.13 & 3.0 & $2.3\times10^{-8}$ & $2.3\times10^{-8}$ & $1.9\times10^{-8}$ & $2.5\times10^{-8}$ \\
1E 2259+586 & 0.059 & 1.3 & $4.6\times10^{-9}$ & $4.7\times10^{-9}$ & $5.1\times10^{-9}$ & $6.0\times10^{-9}$ \\
PSR J1846$-$0258$^\ddagger$ & 0.049 & 1.1 & $3.2\times10^{-9}$ & $3.2\times10^{-9}$ & $3.6\times10^{-9}$ & $4.2\times10^{-9}$ \\
Swift J1822.3$-$1606 & 0.014 & 0.3 & $2.6\times10^{-10}$ & $2.7\times10^{-10}$ & $2.8\times10^{-10}$ & $3.2\times10^{-10}$ \\
SGR 0418+5729 & 0.0061 & 0.1 & $5.0\times10^{-11}$ & $5.0\times10^{-11}$ & $5.1\times10^{-11}$ & $5.9\times10^{-11}$ \\
\hline
\end{tabular}
\caption{Idealized group-delay benchmarks using catalogued magnetar field strengths. The LO, NLO and exact columns assume a constant field over $L=R_M$; the last column integrates the scalar profile in Eq.~(\ref{eq:B_field_model}) along a grazing background ray. These values are not predictions for photons emitted from the magnetar surface.}
\label{tab1}
\end{table}

To isolate the effect of the spatial extent of the field, we use an idealized maximal-birefringence configuration with $\theta_{kB}=\pi/2$ and the dipolar radial scaling
\begin{equation}\label{eq:B_field_model}
B(r) = 
\begin{cases}
B_{0}, & r \leq R_{M}, \\
B_{0}\big( \frac{R_{M}}{r}\big)^{3} & r > R_{M}.
\end{cases}
\end{equation}   
where $B_{0}$ is the surface-field normalization. Equation~(\ref{eq:B_field_model}) models the magnitude rather than the full vector dipole. It therefore provides an upper-envelope propagation benchmark; a general trajectory must retain both $B(r)$ and $\theta_{kB}(r)$. With $n_{\parallel,\perp}=n_{\parallel,\perp}[B(r)]$, the low-energy group delay becomes
\begin{equation}\label{eq:Delta_t_int}
\Delta t = \frac{1}{c}\int_{-\infty}^{\infty} \Delta n_g\big[B(\sqrt{b^2+z^2})\big]\, dz
\end{equation}
where $r =\sqrt{b^2+z^2}$ and $b$ is the impact parameter of the incident light. In our calculation, we set $b = R_{M}$, which corresponds to a grazing trajectory to achieve the maximal impact of the magnetic field of the magnetar. We also assume that light arrives from a far source behind the magnetar, and travels to the (far) detector on Earth. Estimating the time delay using this model requires solving Adler's integral given in Eq.~(\ref{eq:Adler2})-(\ref{eq:Adler4}). However, this raises a complication, namely, Adler's integral was derived for a constant background $B$ field, whereas in our model $B = B(r)$. To circumvent this issue, we use the Locally Constant Field Approximation (LCFA). The LCFA assumes that if the external electromagnetic field varies slowly over the quantum formation time/length of the relevant process, then the field can be treated approximately as constant. In our case, the relevant quantum length is the electron's Compton length $\ell_{e} \sim m_{e}^{-1}$. On the other hand, given our model, for $r > R_{M}$, the field variation scale is given by
\begin{equation}\label{eq:B_variation}
	\ell_{B} \sim \frac{B(r)}{|\nabla B(r)|} \sim \frac{B(r)}{B(r)/r} \sim r,
\end{equation}
and one finds that for a typical size of a magnetar $r \sim R_{M} \sim 10$ km, we have $\ell_{B} \gg \ell_{e}$. In addition, in our calculation we work in the $\omega \rightarrow 0$ limit to remain consistent with the exact and approximate solutions which adopt this limit as well. Thus, we also have $\omega \ll m_{e}$, and the LCFA assumption is justified. The integral can be solved numerically. 
\begin{figure}[t!]
\includegraphics[width=0.6\textwidth]{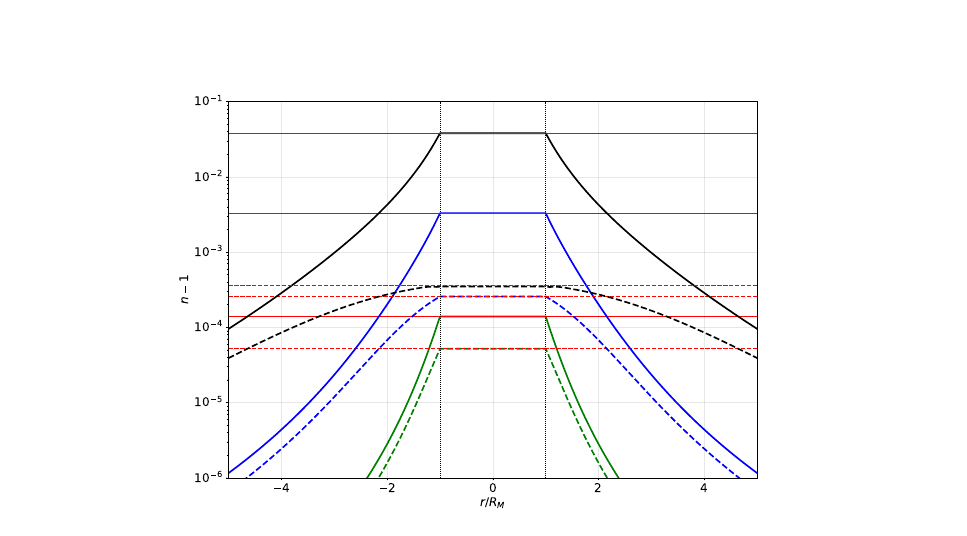}
 \caption{\small{$n_{\parallel}(r)$ (solid) and $n_{\perp}(r)$ (dashed) using the $B$ field model in Eq.~(\ref{eq:B_field_model}) in Adler's integral for $B_{0} =  B_{\text{c}}$, (green), $10 B_{\text{c}}$ (blue), and $100 B_{\text{c}}$ (black) in the limit $\omega \rightarrow 0$. The red lines represent the exact solution obtained from Eqs.~(\ref{eq:exact_sol1})-(\ref{eq:exact_sol4})}}
\label{fig:n(r)} 
\end{figure}

We begin by calculating $n_{\parallel,\perp}(r)$ using Adler's integral in the limit $\omega \rightarrow 0$. We show the results in Figure~\ref{fig:n(r)} for $B_{0}/B_{\text{c}} = 1$, $10$, and $100$. We also plot the exact solution for a constant $B_{0}$ obtained from Eqs.~(\ref{eq:exact_sol1})\,--\,(\ref{eq:exact_sol4}) for reference.\footnote{Note that this solution is only valid for $r\sim R_{M}$. We only extend the lines beyond that for visual clarity.} We find that within $r/R_{M} = \pm 1$ corresponding to the vicinity of the magnetar's surface, our model calculated numerically agrees with the exact solution. Beyond this region, $n_{\parallel,\perp}(r)-1$ drop rapidly but remain non-negligible, which suggests that the contribution of the magnetic field beyond the vicinity of the magnetar's surface should be taken into account.

We calculate $\Delta t$ using Eq.~(\ref{eq:Delta_t_int}) for the catalog field strengths and show the results in the last column of Table~\ref{tab1}. Relative to the exact constant-field estimate with $L=R_M$, the extended profile enhances the delay by a factor of roughly $1.2$--$1.7$ across the table and reaches $0.96\,\mu{\rm s}$ for the largest field. This comparison should not be confused with an end-to-end detectability forecast: resolving such a delay requires a suitably variable background source and an analysis capable of statistically separating the two polarization components.

\begin{figure}[t!]
\centering
\includegraphics[width=0.48\textwidth]{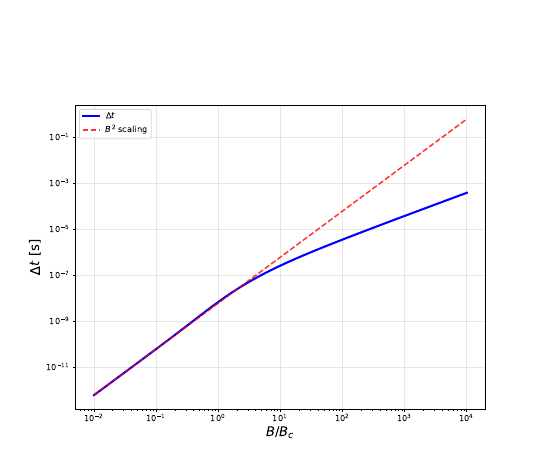}
\hfill
\includegraphics[width=0.48\textwidth]{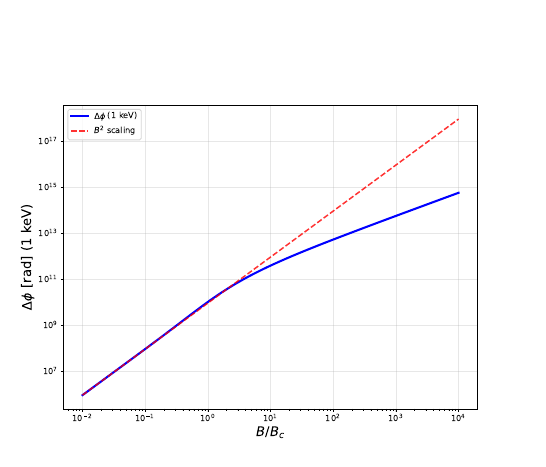}
 \caption{\small{(Left) The time delay $\Delta t$ vs $\frac{B}{B_{\text{c}}}$. (Right) The phase difference $\Delta \phi$ vs $\frac{B}{B_{\text{c}}}$. The red dashed line represents the $\Delta t, \Delta \phi \propto B^{2}$ scaling as implied by Eq.~(\ref{eq:Delta_n}).}}
\label{fig:Delta_t_phi}
\end{figure}

\begin{figure}[t!]
\includegraphics[width=0.6\textwidth]{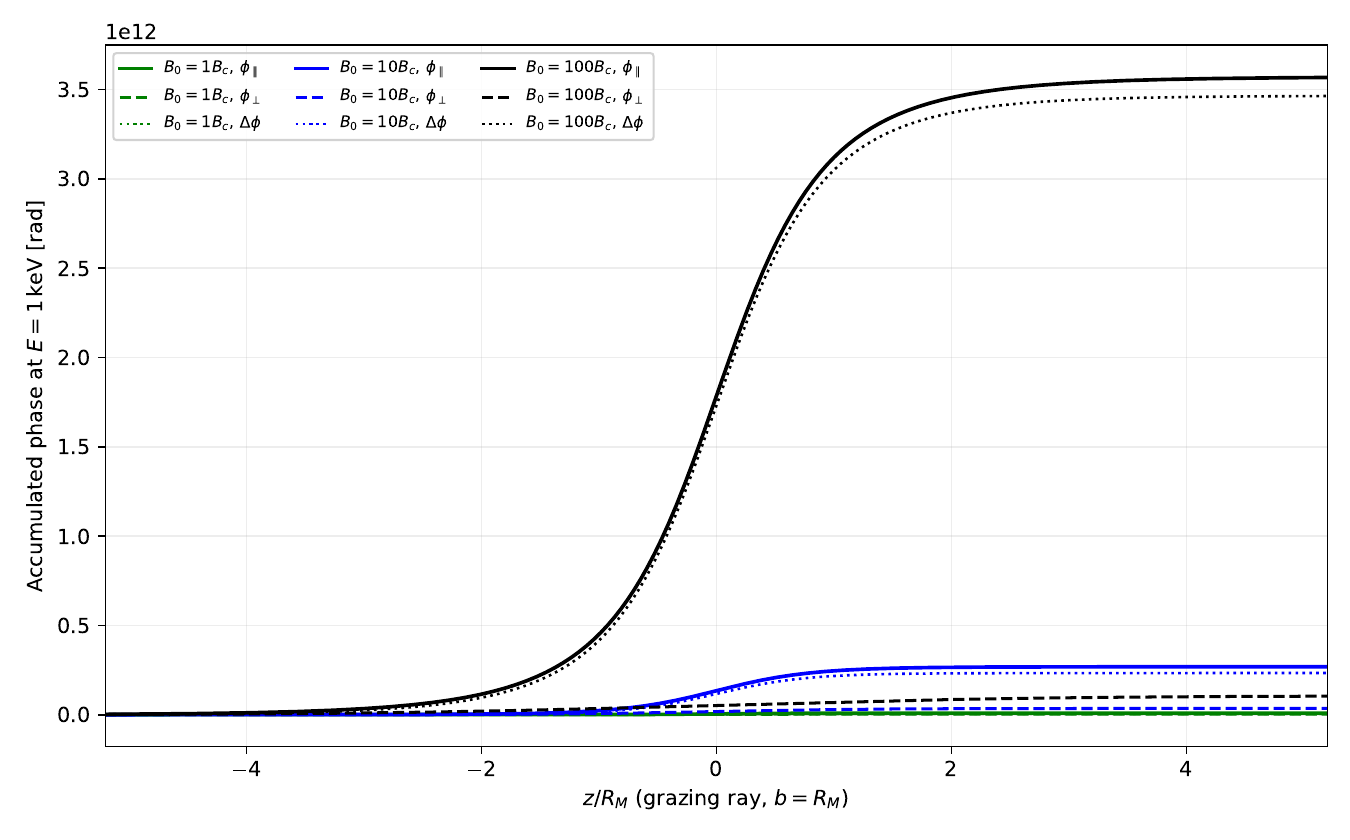}
 \caption{\small{The phase retardation for the parallel mode (dotted) and the perpendicular mode (dashed), and the phase difference between the two modes (dark solid). The source is assumed to be located at $-\infty$ and the observer is assumed to be located at $+\infty$. The photon energy is $E=1$ keV and the ray grazes the stellar surface ($b=R_M$).}}
\label{fig:detla_phi} 
\end{figure}

We plot the time delay $\Delta t$ vs. $B/B_{\text{c}}$ in the left panel of Figure~\ref{fig:Delta_t_phi}. As the plot shows, for $B/B_{\text{c}} \lesssim 1$, the weak field expansion is valid and from Eq.~(\ref{eq:Delta_n}), we see that $\Delta t \propto \Delta n \propto B^{2}$. As $B/B_{\text{c}}$ increases, the scaling deviates significantly from $\Delta t \propto B^{2}$, and we see that $d(\Delta t)/d(B/B_{\text{c}})$ becomes a decreasing function with $B/B_{\text{c}}$. In fact, in the strong field limit $B/B_{\text{c}} \rightarrow \infty$ (corresponding to $\overline{a} \rightarrow 0$), the exact solution in Eqs.~(\ref{eq:exact_sol1})\,--\,(\ref{eq:exact_sol4}) implies that
\begin{align}
n_{\perp}(B/B_{\text{c}}\gg 1) & \approx 1 + \frac{\alpha}{6\pi},\label{eq:n_perp_asymptot}\\
n_{\parallel}(B/B_{\text{c}}\gg 1) & \approx \sqrt{1+\frac{\alpha}{3\pi}\frac{B}{B_{\text{c}}}}\label{eq:n_parallel_asymptot},\\
&\notag
\end{align}
\addtocounter{equation}{1} 
which implies that $\Delta n (B/B_{\text{c}} \gg 3\pi/\alpha) \propto \sqrt{\frac{B}{B_{\text{c}}}}$ thereby explaining the scaling observed in Figure~\ref{fig:Delta_t_phi} for $B/B_{\text{c}} \gg 1$. 
In addition to $\Delta t$, birefringence will lead to a phase difference between the parallel and perpendicular eigenmodes. For an incident photon of energy $\omega$, the accumulated phase retardation for each mode can be found as follows 
\begin{equation}\label{eq:phase_integral}
\Delta \phi_{\parallel,\perp} = \frac{E}{\hbar c}\int_{-\infty}^{\infty}dz\,\left[n_{\parallel,\perp}(B(\sqrt{b^2+z^2}))-1\right],
\end{equation}
with the relative phase difference given by $\Delta \phi = \Delta \phi_{\parallel} - \Delta \phi_{\perp}$. The right panel of Figure~\ref{fig:Delta_t_phi} shows $\Delta \phi$ against $B/B_{\text{c}}$ for $\omega=1$ keV, with scaling similar to the delay. Figure~\ref{fig:detla_phi} shows the accumulated phase along the grazing path. Most of the phase is acquired in the inner magnetosphere and the integral saturates as the field decays. Saturation of this scalar integral is not, by itself, the polarization freeze-out condition. Freeze-out is determined by the competition between the local mode splitting and rotation of the eigenaxes, as derived in Section~\ref{sec4}. The absolute phase is also subject to rapid wrapping and finite-band averaging; the directly relevant quantities are the transported Stokes parameters.

\begin{figure}[b!]
\centering
\begin{subfigure}{0.32\textwidth}
    \centering
    \includegraphics[width=\linewidth]{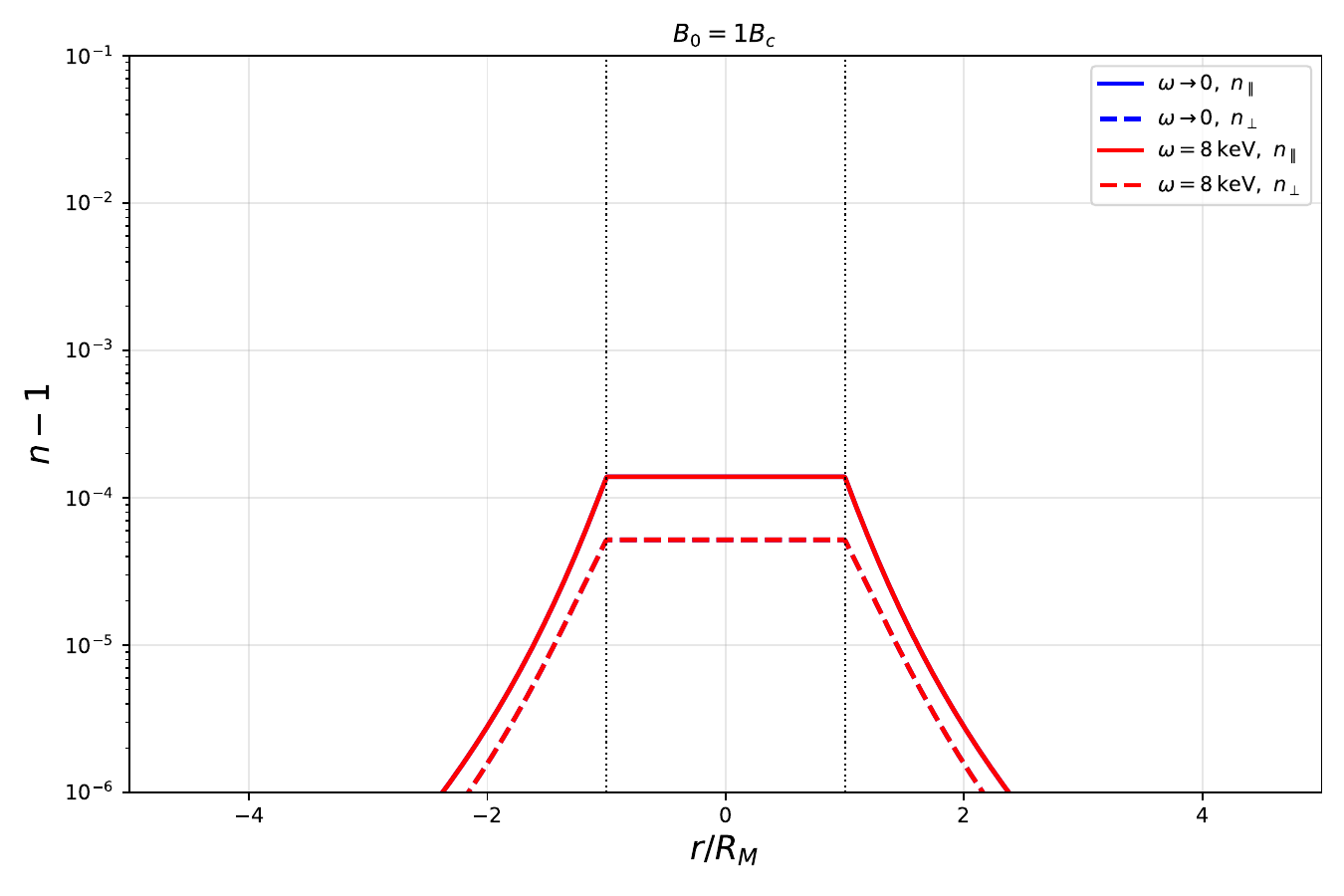}
\end{subfigure}
\hfill
\begin{subfigure}{0.32\textwidth}
    \centering
    \includegraphics[width=\linewidth]{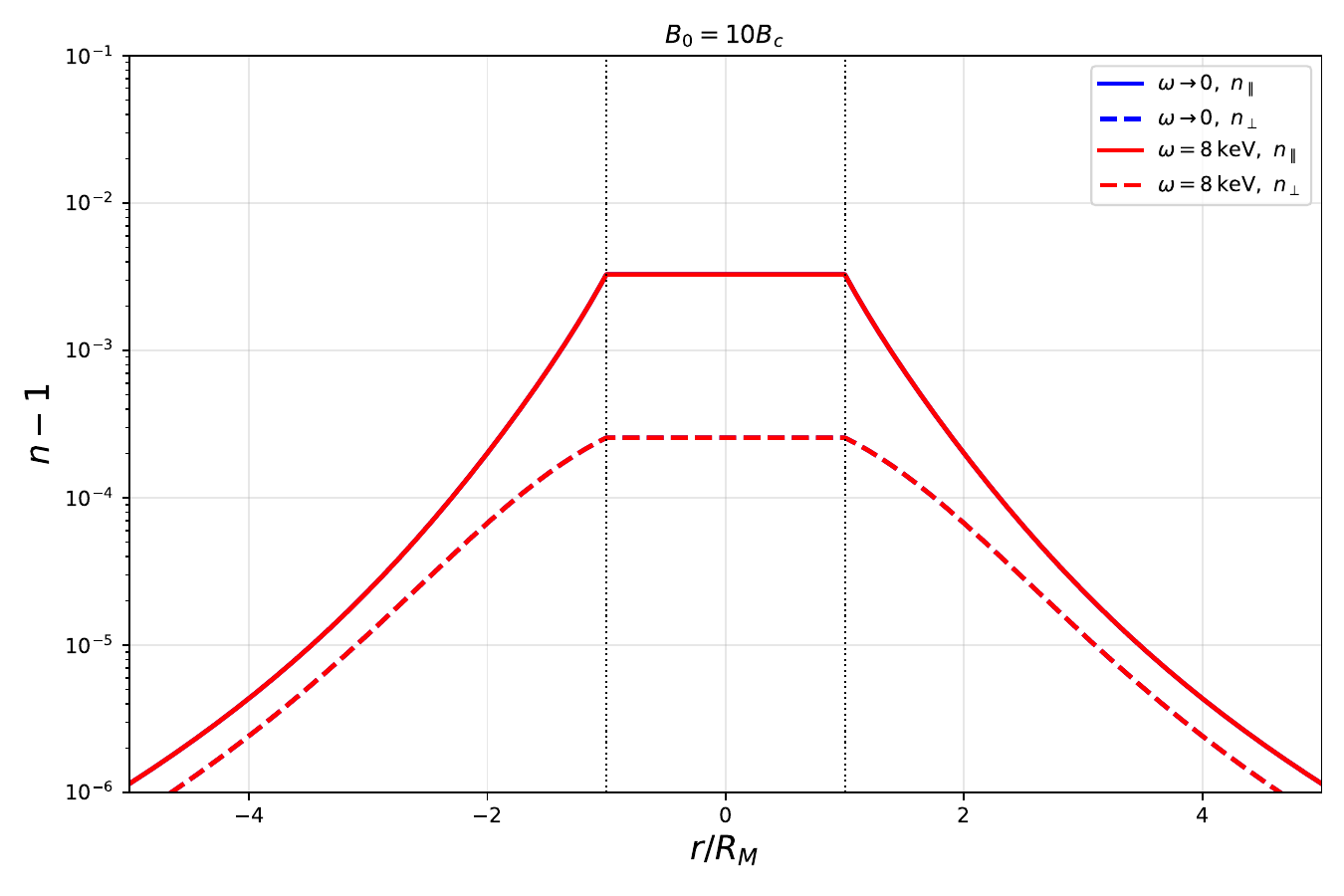}
\end{subfigure}
\hfill
\begin{subfigure}{0.32\textwidth}
    \centering
    \includegraphics[width=\linewidth]{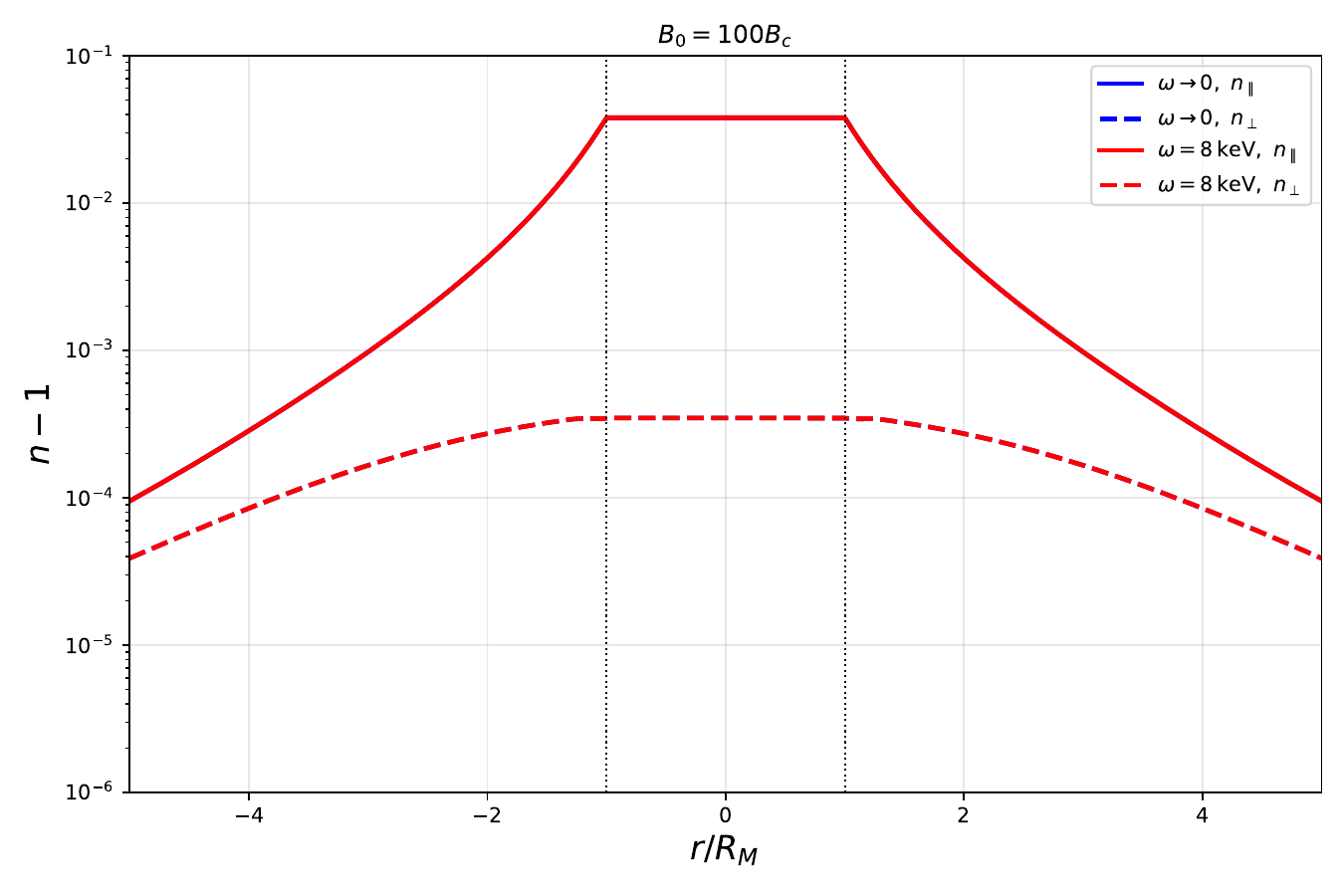}
\end{subfigure}

\caption{\small{Comparison between $n_{\parallel}$ (solid) and $n_{\perp}$ (dashed) for photon energy $\omega = 0$ keV (blue) and $8$ keV (red) using Adler's integral formula for $B/B_{\text{c}}=1$ (left), $10$ (middle), and $100$ (right).}}
\label{fig:omega_comp}
\end{figure}

Before we conclude this section, we comment on the sensitivity of our results to the energy of the incident photon $\omega$. The IXPE experiment is sensitive to energies between $2$–$8~\mathrm{keV}$~\cite{Muleri:2021wpd}, while the Polarimetry Focusing Array (PFA) in the eXTP experiment will be sensitive to the energy range between $2$–$10~\mathrm{keV}$~\cite{eXTP:2018anb}. Thus, it is worth investigating the robustness of our results for energies in this range. It is quite obvious that $\Delta \phi \propto \omega$. However, we find that $\Delta n$ and consequently $\Delta t$ are insensitive to these energies where $ \omega \ll m_{e}$. We show $n_{\parallel,\perp}$ using Adler's integral formula for both $\omega = 0$ and $8~\mathrm{keV}$ for $B/B_{\text{c}} =1$, $10$, and $100$ in Figure~\ref{fig:omega_comp}. As the plot shows, the two cases are essentially indistinguishable from one another. We have verified numerically that for energies up to $\omega = 100$ keV, the solutions agree at the sub-percent level, whereas for $\omega = 300~(500)~\mathrm{keV}$, they are off by $\sim 7~(24)\%$. These energies are well above the sensitivity range of current X-ray polarimeters and thus our results are very robust.

\section{Polarization Transport and Observational Benchmarks}\label{sec4}

\subsection{Evolution in a spatially varying eigenbasis}

IXPE measures linear X-ray polarization in the $2$--$8$ keV band~\cite{Muleri:2021wpd}. The future eXTP mission, currently approved for launch in 2030, will include a larger-area Polarimetry Focusing Array~\cite{eXTP:2025}. Neither instrument measures the accumulated phase in Eq.~(\ref{eq:phase_integral}) directly. Their fundamental observables are constructed from the Stokes parameters $S=(I,Q,U,V)$.

Let $\psi_B(s)$ denote the position angle of the magnetic field projected onto the plane transverse to the ray, measured relative to a fixed sky basis. In a transparent vacuum the intensity is conserved, while the reduced Stokes vector $\boldsymbol{s}=(Q,U,V)$ obeys~\cite{HeylShaviv:1999,HeylShaviv:2002,WangLai:2009} (see the next section for details)
\begin{equation}\label{eq:stokes_transport}
\frac{dI}{ds}=0,\qquad
\frac{d\boldsymbol{s}}{ds}=\boldsymbol{\Omega}(s)\times\boldsymbol{s},\qquad
\boldsymbol{\Omega}(s)=\Delta k(s)\big(\cos 2\psi_B(s),\sin 2\psi_B(s),0\big),
\end{equation}
where
\begin{equation}\label{eq:delta_k}
\Delta k(s)=\frac{\omega}{c}\Delta n\big[B(s),\theta_{kB}(s)\big].
\end{equation}
Equivalently,
\begin{equation}\label{eq:stokes_matrix}
\frac{d}{ds}
\begin{pmatrix}Q\\U\\V\end{pmatrix}
=
\begin{pmatrix}
0&0&\Omega_U\\
0&0&-\Omega_Q\\
-\Omega_U&\Omega_Q&0
\end{pmatrix}
\begin{pmatrix}Q\\U\\V\end{pmatrix}.
\end{equation}
Equation~(\ref{eq:stokes_transport}) preserves $Q^2+U^2+V^2$ for each monochromatic coherent ray. Depolarization arises after averaging rays with different phases, emission locations, magnetic geometries or photon energies; it is not produced by lossless propagation of one ray.

\begin{figure}[t!]
\centering
\includegraphics[width=0.95\linewidth]{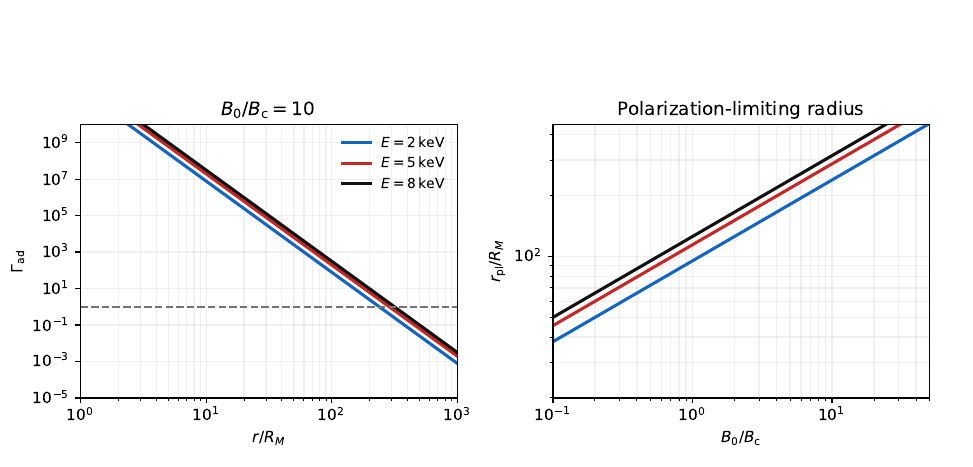}
\caption{\small{Polarization transport in the weak-field outer magnetosphere. Left: $\Gamma_{\rm ad}=(r_{\rm pl}/r)^5$ for $B_0/B_{\rm c}=10$ and representative IXPE energies. The dashed line marks freeze-out. Right: polarization-limiting radius from Eq.~(\ref{eq:rpl}) for $R_M=10$ km and $f_{\rm geom}=1$. Order-one geometric factors shift these curves only through a fifth root.}}
\label{fig:transport}
\end{figure}

When the splitting is large compared with the rate at which the local eigenaxes rotate, the polarization evolves adiabatically and follows the projected magnetic field. A useful adiabaticity measure is
\begin{equation}\label{eq:adiabaticity}
\Gamma_{\rm ad}(s)\equiv\frac{|\Delta k(s)|}{2|d\psi_B/ds|}.
\end{equation}
The evolution is adiabatic for $\Gamma_{\rm ad}\gg1$ and freezes when $\Gamma_{\rm ad}\sim1$, defining the polarization-limiting radius $r_{\rm pl}$. At the large radii relevant to freeze-out, the weak-field expression
\begin{equation}\label{eq:weak_delta_n_transport}
\Delta n\simeq\frac{\alpha}{30\pi}\left(\frac{B_\perp}{B_{\rm c}}\right)^2
\end{equation}
is sufficient. Taking $B_\perp\simeq B_0(R_M/r)^3$ and absorbing order-one orientation and field-rotation factors into $f_{\rm geom}$ gives
\begin{equation}\label{eq:rpl}
\frac{r_{\rm pl}}{R_M}\simeq
\left[
\frac{\alpha}{30\pi}\frac{E R_M}{\hbar c}
\left(\frac{B_0}{B_{\rm c}}\right)^2 f_{\rm geom}
\right]^{1/5}.
\end{equation}
Figure~\ref{fig:transport} displays the resulting scaling for $f_{\rm geom}=1$. In the IXPE band $r_{\rm pl}$ is typically tens to several hundred stellar radii. The local field there is far below $B_{\rm c}$, so retaining exact finite-field indices changes the near-surface phase but has negligible effect on $r_{\rm pl}$, in agreement with Ref.~\cite{Abbassi:2026}.

For photons emitted by the magnetar, Eq.~(\ref{eq:stokes_transport}) must be supplied with the emergent Stokes vector from the atmosphere or condensed surface. In an atmosphere, plasma birefringence, absorption and emission lead instead to a polarized transfer equation of the form $dS/ds=j-\mathsf{K}S$, and the plasma--vacuum resonance can convert the normal modes~\cite{LaiHo:2003,HoLai:2003,vanAdelsbergLai:2006,Kelly:2024}. Resonant Compton scattering may subsequently modify the signal in a twisted magnetosphere~\cite{FernandezDavis:2011,Taverna:2014,GuoLai:2026}. These processes, together with surface integration, general-relativistic ray bending and viewing geometry, are required for a source-specific IXPE fit and are not included in the catalog benchmark below.

\subsection{Fixed-eigenaxis limit}
The scalar phase mapping used in the remainder of this section is the exactly solvable limit of Eq.~(\ref{eq:stokes_transport}) in which $\psi_B$ is constant, $\theta_{kB}=\pi/2$, and the incident Stokes vector is specified. It is therefore an idealized propagation and instrument-response benchmark, not a prediction for the thermal radiation emitted by each listed magnetar.

Experiments like IXPE and eXTP do not measure $\Delta t$ or $\Delta \phi$ directly, instead, they measure the aforementioned Stokes parameters. The Stokes parameters are a set of four quantities that fully describe the polarization state of electromagnetic radiation. These quantities are the intensity $I$, the horizontal vs. vertical linear polarization $Q$ defined as the difference between the intensities in the horizontal and vertical direction ($Q = I_{H} - I_{V}$), the diagonal linear polarization $U$ defined as the difference in the intensities at $\pm 45\degree$ ($U = I_{45\degree} - I_{135\degree}$), and the circular polarization $V$ defined as the difference in the intensities between the left-handed and right-handed polarizations ($V = I_{R} - I_{L}$). These quantities form the Stokes vector $S = (I, Q, U, V)$. In practice, experiments usually report the following derived quantities
\begin{align}
P & = \frac{\sqrt{Q^{2}+U^{2}+V^{2}}}{I},\label{eq:P}\\
P_{L} & = \frac{\sqrt{Q^{2}+U^{2}}}{I},\label{eq:PL}\\
\chi & = \frac{1}{2}\text{atan2}(U,Q) \hspace{5mm}, \label{eq:chi}
\end{align}
where $P$, $P_{L}$ and $\chi$ are the degree of polarization, linear polarization fraction, and polarization angle, respectively and $\text{atan2}(y,x)$ measures the angle of the vector $(y,x)$ with the positive $x$-axis. IXPE and eXTP are not sensitive to circular polarization $V$ and thus cannot measure $P$, but they can measure $P_{L}$ and $\chi$.

\begin{figure}[t!]
\centering
\includegraphics[width=\linewidth]{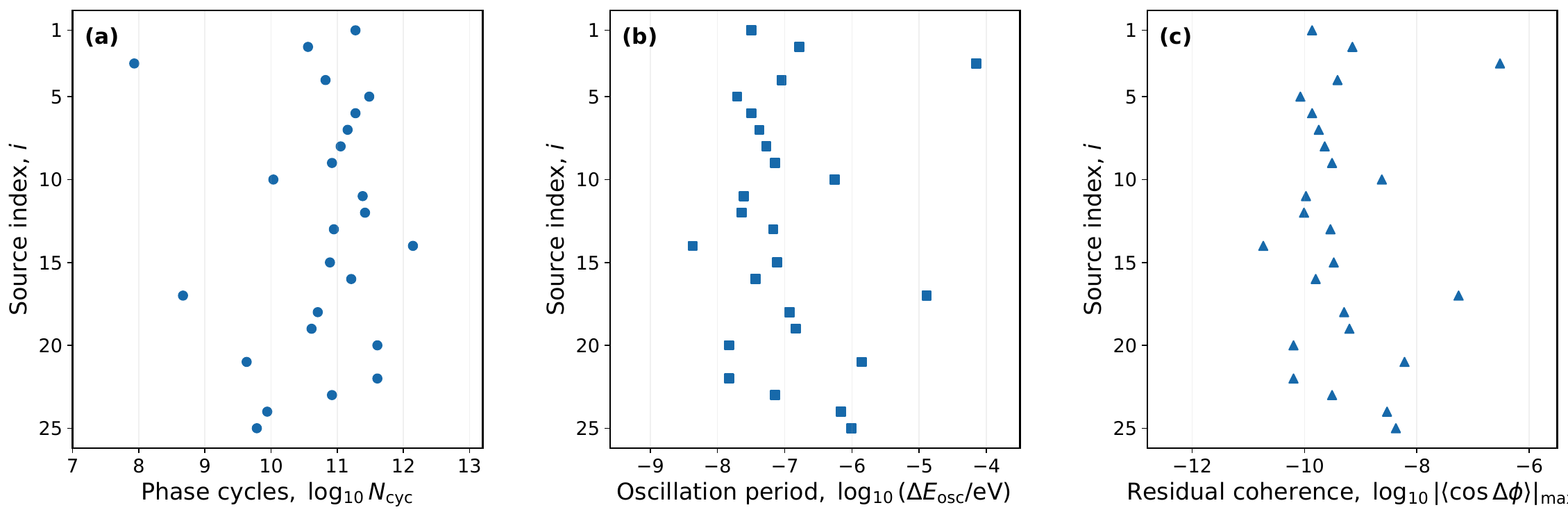}
\caption{\small{Fixed-eigenaxis diagnostics for the 25 catalog magnetic-field benchmarks: (a) the number of phase oscillations across the $2$--$8$ keV band; (b) the corresponding oscillation period in energy; and (c) a conservative upper bound on the magnitude of residual phase coherence in any $0.15$-keV bin, for the illustrative IXPE photon weight $W(E)=S(E)A_{\rm eff}(E)$. Together these panels diagnose the rapid phase oscillation responsible for the saturated band-averaged limit.}}
\label{fig:heatmaps}
\end{figure}

To understand how the measured observables $\chi$ and $P_{L}$ relate to the predictions, let us decompose the propagating electric field into its parallel and perpendicular components $E = E_{\parallel} + E_{\perp}$. After propagating near the strong magnetic field of the magnetar, each component will acquire a phase $E_{\parallel} \rightarrow E_{\parallel} e^{i\Delta\phi_{\parallel}}$ and $E_{\perp} \rightarrow E_{\perp} e^{i\Delta\phi_{\perp}}$, such that the relative phase between the two components is given by $E = E_{\parallel} + E_{\perp} e^{i\Delta \phi}$. From the definitions of the evolution of the Stokes parameters, one finds that they can be expressed as 
\begin{align}
Q_{0} & = |E_{\parallel}|^{2} - |E_{\perp}|^{2},\label{eq:Q}\\
U_{0} & = 2\text{Re}(E_{\parallel}E_{\perp}^{*}),\label{eq:U}\\
V_{0} & = 2\text{Im}(E_{\parallel}E_{\perp}^{*}), \label{eq:V}
\end{align}
which implies that after acquiring the relative phase, these parameters evolve as
\begin{align}
Q & = Q_{0},\label{eq:Q_E}\\
U & = U_{0}\cos{(\Delta \phi)} - V_{0}\sin{(\Delta \phi)} ,\label{eq:U_E}\\
V & = U_{0}\sin{(\Delta \phi)} + V_{0}\cos{(\Delta \phi)},\label{eq:V_E}
\end{align}
where $\chi_{0}$ is the initial orientation angle of the incident photon. Thus, the phase difference $\Delta \phi$ is encoded in the Stokes parameters. Specifically, let us assume as a benchmark that the incident photon is linearly polarized such that $V_{0} = 0$, $P_{L,0} \equiv P_{0} = 1$. Thus, initially one has
\begin{align}
Q_{0} = I_{0} \cos{(2\chi_{0})}, \hspace{5mm} U_{0} = I_{0} \sin{(2\chi_{0})},
\end{align}
and after evolution one has
\begin{align}
P_{L}(E) & = \sqrt{\cos^{2}(2\chi_{0})+\sin^{2}(2\chi_{0})\cos^{2}(\Delta \phi(E))}, \label{eq:PL_E}\\
\chi(E) & = \frac{1}{2}\text{atan2}\Big(\sin(2\chi_{0})\cos(\Delta\phi(E)), \cos(2\chi_{0}) \Big).
\label{eq:chi_E}
\end{align}

To illustrate the strong energy dependence of the phase and its averaging, Figure~\ref{fig:heatmaps} shows the source-dependent quantities relevant to the fixed-eigenaxis approximation. These include the number of phase oscillations across the observing band, the corresponding oscillation period in energy, and an upper bound on the residual phase coherence in a $0.15$-keV bin. The latter is obtained by applying the integration-by-parts bound derived below over each energy bin and taking the largest value. The figure demonstrates that the accumulated phase varies extremely rapidly over the energy range relevant to IXPE and eXTP.

Although $P_L(E)$ and $\chi(E)$ represent physical polarization observables, detectors do not measure these monochromatic quantities with infinite energy resolution. Instead, they measure energy-averaged Stokes parameters, which we express as
\begin{equation}\label{eq:smearedQUI}
\overline{I} = \int dE\,W(E)I(E), \hspace{5mm}
\overline{Q} = \int dE\,W(E)Q(E), \hspace{5mm}
\overline{U} = \int dE\,W(E)U(E),
\end{equation}
where $W(E)$ is an energy-dependent weighting function encoding the source spectrum and detector response. We describe the illustrative weighting functions adopted for IXPE and eXTP in Appendix~\ref{appendixA}.\footnote{The weighting functions used here are simulated approximations; experiment-specific response functions are required for precision observational analyses.} The corresponding observables are given by
\begin{equation}\label{eq:smeared_P_Chi}
\overline{P}_L =
\frac{\sqrt{\overline{Q}^{\,2}+\overline{U}^{\,2}}}{\overline{I}},
\hspace{5mm}
\overline{\chi} =
\frac{1}{2}\operatorname{atan2}(\overline{U},\overline{Q}).
\end{equation}
For the initially fully linearly polarized beam considered here, one obtains
\begin{equation}
\frac{\overline{Q}}{\overline{I}}=\cos(2\chi_0),
\hspace{5mm}
\frac{\overline{U}}{\overline{I}}=\sin(2\chi_0)C_W,
\end{equation}
where
\begin{equation}\label{eq:cos_average}
C_W =
\frac{\displaystyle\int_{2\,\mathrm{keV}}^{8\,\mathrm{keV}}
dE\,W(E)\cos[\Delta\phi(E)]}
{\displaystyle\int_{2\,\mathrm{keV}}^{8\,\mathrm{keV}}dE\,W(E)}.
\end{equation}
In the nondispersive approximation, the phase difference takes the form
\begin{equation}
\Delta\phi(E)=\frac{E\Delta t_0}{\hbar},
\end{equation}
where $\Delta t_0$ is the differential propagation delay. The corresponding energy oscillation period is
\begin{equation}
\Delta E_{\rm osc}=\frac{2\pi\hbar}{\Delta t_0}.
\end{equation}
Even for the smallest constant-field delay in Table~\ref{tab1}, $\Delta t_0\simeq5.0\times10^{-11}$~s, the oscillation period is approximately $8.27\times10^{-8}$~keV, corresponding to more than $7\times10^7$ oscillations across the $2$--$8$ keV band. Therefore, the oscillatory contribution cannot be reliably evaluated using a coarse energy grid.

For the smooth illustrative weighting functions adopted in Appendix~\ref{appendixA}, integration by parts gives the conservative bound
\begin{equation}\label{eq:C_bound}
|C_W|\leq
\frac{\hbar\left[
|W(E_{\min})|+|W(E_{\max})|
+\displaystyle\int_{E_{\min}}^{E_{\max}}dE\,|W'(E)|
\right]}
{\Delta t_0\displaystyle\int_{E_{\min}}^{E_{\max}}dE\,W(E)},
\end{equation}
where energies and $\hbar$ are expressed in consistent units. Using the conservative value $\Delta t_0=5.0\times10^{-11}$~s, we obtain $|C_W|<1.50\times10^{-8}$ for IXPE and $|C_W|<1.08\times10^{-8}$ for eXTP with $W(E)=S(E)A_{\rm eff}(E)$.

Consequently, the oscillatory contribution to $\overline U$ becomes negligible, whereas $\overline Q$ remains unchanged. The resulting band-averaged linear polarization fraction is therefore
\begin{equation}\label{eq:saturatedPL}
\boxed{\overline P_L\simeq|\cos(2\chi_0)|}.
\end{equation}
This result is independent of the magnetar magnetic-field strength in the saturated-phase limit, provided the same initial polarization state and the fixed-eigenaxis approximation are adopted. We summarize the corrected results for IXPE and eXTP in Table~\ref{tab2}, considering the two benchmark initial orientations $\chi_0=45^\circ$ and $30^\circ$.

\begin{table}[t!]
\centering
\begin{tabular}{lcccc}
\hline\hline
Instrument & $\chi_0$ &
$\overline Q/\overline I$ &
$|\overline U/\overline I|$ &
$\overline P_L$ \\
\hline
IXPE & $45^\circ$ & $0$ & $<1.50\times10^{-8}$ & $\simeq0$ \\
IXPE & $30^\circ$ & $0.50$ & $<1.30\times10^{-8}$ & $\simeq0.50$ \\
eXTP & $45^\circ$ & $0$ & $<1.08\times10^{-8}$ & $\simeq0$ \\
eXTP & $30^\circ$ & $0.50$ & $<9.31\times10^{-9}$ & $\simeq0.50$ \\
\hline\hline
\end{tabular}
\caption{\small{Fixed-eigenaxis band-averaged polarization benchmarks for IXPE and eXTP, assuming an initially fully linearly polarized beam. The bounds on the residual $\overline U/\overline I$ are obtained using the illustrative photon-count weighting functions of Appendix~\ref{appendixA} and the nondispersive approximation. The results are common to the catalog magnetic-field benchmarks in the saturated-phase limit.}}
\label{tab2}
\end{table}

We see that for $\chi_0=45^\circ$, the band-averaged linear polarization is strongly suppressed, $\overline P_L\simeq0$, and the polarization angle becomes undefined in the saturated limit. On the other hand, for $\chi_0=30^\circ$, the surviving linear polarization fraction approaches $\overline P_L\simeq0.50$, with $\overline\chi\simeq0^\circ$. These results are consistent with the disappearance of the rapidly oscillating contribution to the Stokes parameter $U$ after energy averaging.

An important consequence is that the corrected polarization fractions are essentially independent of the magnetar magnetic-field strength within the present approximation. Although stronger magnetic fields lead to larger accumulated phase differences and shorter oscillation periods, these differences do not translate into distinct band-averaged polarization fractions once the rapid-phase regime is reached. Thus, the fixed-eigenaxis calculation does not provide a basis for ranking individual magnetars according to their polarization detectability.

For completeness, we perform an illustrative signal-to-noise ratio (SNR) analysis for measuring the surviving linear polarization with IXPE and eXTP. As discussed in Appendix~\ref{appendixB}, the relevant sensitivity depends on the source count rate, instrumental modulation factor, background count rate and exposure time. We adopt a representative $2$--$8$ keV energy flux $F_X=2\times10^{-11}\,\mathrm{erg}\,\mathrm{cm}^{-2}\,\mathrm{s}^{-1}$, a background rate $R_B=0.005\,\mathrm{s}^{-1}$, and an exposure $T=100\,\mathrm{ks}$, together with the illustrative instrument-response functions specified in Appendix~\ref{appendixA}.

\begin{figure}[t!]
\centering
\includegraphics[width=0.7\linewidth]{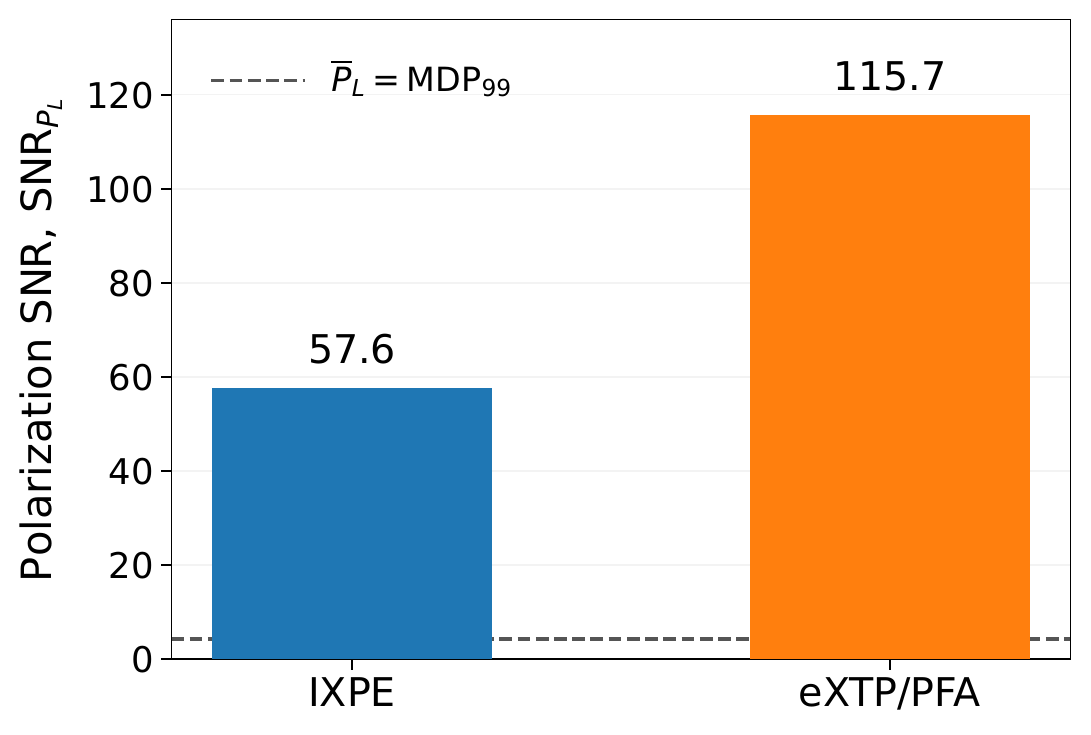}
\caption{\small{Illustrative SNR for measuring the residual band-averaged linear polarization $\overline P_L$ with IXPE and eXTP, assuming $\chi_0=30^\circ$. The calculation adopts the same $2$--$8$ keV source flux $F_X=2\times10^{-11}\,\mathrm{erg\,cm^{-2}\,s^{-1}}$, exposure time $T=100\,\mathrm{ks}$, and background count rate $R_B=0.005\,\mathrm{s^{-1}}$ for both instruments. The horizontal dashed line corresponds to $\mathrm{SNR}_{P_L}=4.292$, equivalent to $\overline P_L=\mathrm{MDP}_{99}$. The SNR quantifies the sensitivity to nonzero linear polarization.}}
\label{fig:SNR}
\end{figure}

We focus on the $\chi_0=30^\circ$ benchmark, for which the residual linear polarization remains appreciable. For this case, we obtain $\mathrm{SNR}_{P_L}\simeq57.6$ for IXPE and $\mathrm{SNR}_{P_L}\simeq115.7$ for eXTP, as shown in Figure~\ref{fig:SNR}. The larger SNR for eXTP follows from its larger assumed effective area and the corresponding increase in the source count rate. The horizontal dashed line corresponds to $\mathrm{SNR}_{P_L}=4.292$, or equivalently $\overline P_L=\mathrm{MDP}_{99}$, under the convention adopted in Appendix~\ref{appendixB}.

We emphasize that the SNR calculated here characterizes the statistical sensitivity to nonzero linear polarization, rather than the significance of a direct detection of QED vacuum birefringence. In particular, the incident polarization state has been prescribed, and the same illustrative flux has been adopted for the magnetar benchmarks. A quantitative determination of birefringence detectability requires a comparison between source-specific models with and without vacuum birefringence, including the emission mechanism, viewing geometry, plasma effects, and magnetospheric scattering. Such an analysis lies beyond the scope of the present study.

\section{Conclusions}\label{sec5}
We have studied finite-field one-loop vacuum birefringence along an idealized grazing ray from a distant background source passing a magnetar. Using a spatially decaying dipole-magnitude profile instead of a constant field truncated after one stellar radius enhances the finite-field group delay by a factor of roughly $1.2$--$1.7$ for the catalog field strengths. The largest benchmark delay is approximately $0.96\,\mu{\rm s}$. The calculation is an upper-envelope propagation estimate because it assumes $\theta_{kB}=\pi/2$, a straight ray and a scalar field profile; a general result requires the full vector field, gravitational and magnetic ray bending, and a specified background transient or periodic source.

We have also supplied the polarization-state evolution in a varying eigenbasis. In a varying projected magnetic field the reduced Stokes vector precesses about the local birefringence vector according to Eq.~(\ref{eq:stokes_transport}). The modes follow the magnetic direction adiabatically until the splitting becomes comparable to the rotation rate of the eigenaxes, at which point the polarization freezes. For X-ray photons this occurs far enough from the star that $B\ll B_{\rm c}$, explaining why exact finite-field corrections can substantially affect the near-surface accumulated phase while leaving the polarization-limiting radius essentially unchanged.

The band averages and SNR values have a deliberately restricted interpretation. They solve the fixed-eigenaxis limit for a prescribed, fully linearly polarized input and common illustrative detector weights. The oscillatory Stokes contribution averages out across the instrument bands, so $\overline{P}_L\simeq|\cos(2\chi_0)|$ and the original source-specific ranking is not supported. The SNR plot concerns only the measurement of nonzero linear polarization in the representative $\chi_0=30^\circ$ benchmark, not a likelihood-ratio detection test of vacuum birefringence. A source-specific analysis must combine atmosphere or condensed-surface emission, plasma--vacuum mode conversion, surface and viewing geometry, general-relativistic ray tracing and, when relevant, magnetospheric scattering. The exact finite-field indices calculated here can be inserted into such a transport pipeline; the plasma--vacuum resonance is a particularly promising place to test their observational effect~\cite{Abbassi:2026}.

\section*{Acknowledgment}

We thank Sudhir Vempati and Nirupam Roy for the valuable discussions in the initial stages of this project. \textit{AI Usage Declaration:} The author acknowledges the use of OpenAI's ChatGPT for assistance with literature organization, manuscript drafting, code writign and language editing. The author assumes full responsibility for the scientific content, interpretations, and references presented in this review.

\appendix
\section{Simulated Weight Functions}\label{appendixA}
To account for the energy dependence of the source and instrument, the physical Stokes parameters are integrated with an illustrative photon-count weight. For both IXPE and eXTP, we use
\begin{equation}\label{eq:W(E)}
W(E) = N S(E)A_{\text{eff}}(E),
\end{equation}
where $S(E)$ is the photon spectrum, $A_{\text{eff}}(E)$ is the effective area, and $N$ is a normalization factor that cancels in the polarization ratios. The modulation factor $\mu(E)$ expresses the instrument's sensitivity to polarization and enters the SNR calculation in Appendix~\ref{appendixB}, not the physical Stokes weighting. For IXPE, we model these functions as
\begin{align}
S(E) & = E^{-\gamma}, \hspace{5mm}\\
A_{\text{eff}} & = \begin{cases}
0, & E \approx 2~\text{keV} \\
500~\text{cm}^{2}, & E \approx 2.3~\text{keV} \\
525~\text{cm}^{2}, & E \approx 3.0~\text{keV} \\
525~\text{cm}^{2}, & E \approx 6.0~\text{keV} \\
0, & E \approx 8 ~\text{keV} \\
\end{cases}\\
\mu(E) & = 0.2 + \Big(\frac{0.55 -0.20}{8-2}\Big)\times(E-2),
\end{align}
while for eXTP, we model them as
\begin{align}
S(E) & = E^{-\gamma}, \hspace{5mm}\\
A_{\text{eff}} & = \begin{cases}
700~\text{cm}^{2}, & E \approx 2.0~\text{keV} \\
1200~\text{cm}^{2}, & E \approx 3.0~\text{keV} \\
1500~\text{cm}^{2}, & E \approx 6.0~\text{keV} \\
1200~\text{cm}^{2}, & E \approx 8.0~\text{keV} \\
\end{cases}\\
\mu(E)  & = \begin{cases}
0.25, & E \approx 2.0~\text{keV} \\
0.40, & E \approx 4.0~\text{keV} \\
0.55, & E \approx 8.0~\text{keV} \\
\end{cases}.
\end{align}
where we adopt a representative photon index $\gamma = 2$ for X-rays, and these values are selected to best simulate the IXPE and eXTP detectors. Notice here that for both $A_{\text{eff}}$ and $\mu(E)$, the intermediate values are interpolated. Actual calibration data once available should be used for precision measurements.

\section{Illustrative Polarization SNR}\label{appendixB}
A commonly used sensitivity metric in X-ray polarimetry is the $99\%$ Minimum Detectable Polarization ($\text{MDP}_{99}$), which represents the smallest polarization fraction detectable at $99\%$ confidence under the adopted counting assumptions. It can be approximated by
\begin{equation}
\text{MDP}_{99} \approx \frac{4.292}{\mu_{\text{eff}} R_{S}}\sqrt{\frac{R_{S}+R_{B}}{T}},
\end{equation}
where $R_{S}$ is the source count rate, $R_{B}$ is the background count rate, $T$ is the exposure time, and $\mu_{\text{eff}}$ is the count-weighted modulation factor. $R_{S}$ and $\mu_{\text{eff}}$ are given by
\begin{align}
R_{S} & = \frac{N_{S}}{T} = \int S(E) A_{\text{eff}}(E)\, dE,\\
\mu_{\text{eff}} & =\frac{\int S(E) A_{\text{eff}}(E)\mu(E)dE}{\int S(E) A_{\text{eff}}(E)\, dE}.
\end{align}
where $S(E)$ is normalized to the assumed source flux. Thus, an illustrative polarization-measurement SNR is
\begin{equation}\label{eq:PL_NSR}
\text{SNR}_{P_{L}} \approx \frac{4.292 \overline{P}_{L}}{\text{MDP}_{99}}.
\end{equation}
The threshold $\overline{P}_L=\mathrm{MDP}_{99}$ corresponds to $\mathrm{SNR}_{P_L}=4.292$, not unity. This statistic tests nonzero polarization, not vacuum birefringence. A birefringence-detection statistic requires explicit with-vacuum and without-vacuum source models convolved with the calibrated response and compared in a common likelihood.


\begin{thebibliography}{10}
\bibitem{Heisenberg:1936nmg}
W.~Heisenberg and H.~Euler,
``Consequences of Dirac's theory of positrons,''
Z. Phys. \textbf{98}, no.11-12, 714-732 (1936)
\arXivold{physics/0605038}{physics}.


\bibitem{Weisskopf:1936}
V. F. Weisskopf, Dan. Mat. Fys. Medd. 14, 1 (1936)


\bibitem{Schwinger:1951nm}
J.~S.~Schwinger,
``On gauge invariance and vacuum polarization,''
Phys. Rev. \textbf{82}, 664-679 (1951)


\bibitem{Meitner:1933kww}
L.~Meitner and H.~K{\"o}sters,
``{\"U}ber die Streuung kurzwelliger {\ensuremath{\gamma}}-Strahlen,''
Z. Phys. \textbf{84}, no.3-4, 137-144 (1933)

\bibitem{Delbruck:1933pla}
M.~Delbr{\"u}ck,
``Note added in proof by M. Delbr{\"u}ck,''
Z. Phys. \textbf{84}, no.3-4, 144 (1933)


\bibitem{Sauter:1931zz}
F.~Sauter,
``Uber das Verhalten eines Elektrons im homogenen elektrischen Feld nach der relativistischen Theorie Diracs,''
Z. Phys. \textbf{69}, 742-764 (1931)


\bibitem{Adler:1971wn}
S.~L.~Adler,
``Photon splitting and photon dispersion in a strong magnetic field,''
Annals Phys. \textbf{67}, 599-647 (1971)



\bibitem{Moreh:1973gma}
R.~Moreh and S.~Kahana,
``Delbruck scattering of 7.9 MeV photons,''
Phys. Lett. B \textbf{47}, 351-354 (1973)


\bibitem{Rullhusen:1983zz}
P.~Rullhusen, U.~Zurmuhl, F.~Smend, M.~Schumacher, H.~G.~Borner and S.~A.~Kerr,
``Giant dipole resonance and Coulomb correction effect in Delbruck scattering studied by elastic and Raman scattering of 8.5 to 11.4 MeV photons,''
Phys. Rev. C \textbf{27}, 559-568 (1983)

\bibitem{Jarlskog:1973aui}
G.~Jarlskog, L.~Joensson, S.~Pruenster, H.~D.~Schulz, H.~J.~Willutzki and G.~G.~Winter,
``Measurement of delbrueck scattering and observation of photon splitting at high energies,''
Phys. Rev. D \textbf{8}, 3813-3823 (1973)

\bibitem{dEnterria:2013zqi}
D.~d'Enterria and G.~G.~da Silveira,
``Observing light-by-light scattering at the Large Hadron Collider,''
Phys. Rev. Lett. \textbf{111}, 080405 (2013)
[erratum: Phys. Rev. Lett. \textbf{116}, no.12, 129901 (2016)]
\arXivold{1305.7142}{hep-ph}.


\bibitem{ATLAS:2017fur}
M.~Aaboud \textit{et al.} [ATLAS],
``Evidence for light-by-light scattering in heavy-ion collisions with the ATLAS detector at the LHC,''
Nature Phys. \textbf{13}, no.9, 852-858 (2017)
\arXivold{1702.01625}{hep-ex}.

\bibitem{CMS:2018erd}
A.~M.~Sirunyan \textit{et al.} [CMS],
``Evidence for light-by-light scattering and searches for axion-like particles in ultraperipheral PbPb collisions at $\sqrt{s_\mathrm{NN}} =$ 5.02 TeV,''
\arXivold{1810.04602}{hep-ex}.

\bibitem{Ellis:2022uxv}
J.~Ellis, N.~E.~Mavromatos, P.~Roloff and T.~You,
``Light-by-light scattering at future $e^+e^-$ colliders,''
Eur. Phys. J. C \textbf{82}, no.7, 634 (2022)
\arXivold{2203.17111}{hep-ph}.

\bibitem{Yang:2020rjt}
J.~C.~Yang, Z.~B.~Qing, X.~Y.~Han, Y.~C.~Guo and T.~Li,
``Tri-photon at muon collider: a new process to probe the anomalous quartic gauge couplings,''
JHEP \textbf{22}, 053 (2020)
\arXivold{2204.08195}{hep-ph}.

\bibitem{Amarkhail:2023xsc}
H.~Amarkhail, S.~C.~Inan and A.~V.~Kisselev,
``Probing anomalous {\ensuremath{\gamma}}{\ensuremath{\gamma}}{\ensuremath{\gamma}}{\ensuremath{\gamma}} couplings at a future muon collider,''
Nucl. Phys. B \textbf{1005}, 116592 (2024)
\arXivold{2306.03653}{hep-ph}.

\bibitem{Spor:2024nsx}
S.~Spor and E.~Gurkanli,
``Analysis of anomalous $H\gamma\gamma$ coupling in light-by-light collision at future muon collider,''
\arXivold{2412.02346}{hep-ph}.

\bibitem{Yoon:2021ony}
J.~W.~Yoon, J.~W.~Yoon, Y.~G.~Kim, Y.~G.~Kim, I.~W.~Choi, I.~W.~Choi, J.~H.~Sung, J.~H.~Sung, H.~W.~Lee and S.~K.~Lee, \textit{et al.}
``Realization of laser intensity over 1023{\,}W/cm2,''
Optica \textbf{8}, no.5, 630-635 (2021)

\bibitem{Danson:2019qlu}
C.~N.~Danson, C.~Haefner, J.~Bromage, T.~Butcher, J.~C.~F.~Chanteloup, E.~A.~Chowdhury, A.~Galvanauskas, L.~A.~Gizzi, J.~Hein and D.~I.~Hillier, \textit{et al.}
``Petawatt and exawatt class lasers worldwide,''
High Power Laser Sci. Eng. \textbf{7}, e54 (2019)

\bibitem{DellaValle:2015xxa}
F.~Della Valle, A.~Ejlli, U.~Gastaldi, G.~Messineo, E.~Milotti, R.~Pengo, G.~Ruoso and G.~Zavattini,
``The PVLAS experiment: measuring vacuum magnetic birefringence and dichroism with a birefringent Fabry{\textendash}Perot cavity,''
Eur. Phys. J. C \textbf{76}, no.1, 24 (2016)
\arXivold{1510.08052}{physics.optics}.


\bibitem{Taverna:2022jgl}
R.~Taverna, R.~Turolla, F.~Muleri, J.~Heyl, S.~Zane, L.~Baldini, D.~G.~Caniulef, M.~Bachetti, J.~Rankin and I.~Caiazzo, \textit{et al.}
``Polarized x-rays from a magnetar,''
\arXivold{2205.08898}{astro-ph.HE}.

\bibitem{McGill:2026}
``McGill Online Magnetar Catalog,''
\url{http://www.physics.mcgill.ca/~pulsar/magnetar/main.html}


\bibitem{Born:1934gh}
M.~Born and L.~Infeld,
``Foundations of the new field theory,''
Proc. Roy. Soc. Lond. A \textbf{144}, no.852, 425-451 (1934)


\bibitem{Biswas:2014yia}
T.~Biswas and N.~Okada,
``Towards LHC physics with nonlocal Standard Model,''
Nucl. Phys. B \textbf{898}, 113-131 (2015)
\arXivold{1407.3331}{hep-ph}.


\bibitem{Abu-Ajamieh:2023syy}
F.~Abu-Ajamieh and S.~K.~Vempati,
``A proposed renormalization scheme for non-local QFTs and application to the hierarchy problem,''
Eur. Phys. J. C \textbf{83}, no.11, 1070 (2023)
\arXivold{2304.07965}{hep-th}.

\bibitem{Abu-Ajamieh:2023roj}
F.~Abu-Ajamieh, P.~Chattopadhyay, A.~Ghoshal and N.~Okada,
``Anomalies in string-inspired nonlocal extensions of QED,''
Phys. Rev. D \textbf{109}, no.7, 076013 (2024)
\arXivold{2307.01589}{hep-th}.




\bibitem{Abu-Ajamieh:2023txh}
F.~Abu-Ajamieh, N.~Okada and S.~K.~Vempati,
``Corrected calculation for the non-local solution to the g {\ensuremath{-}} 2 anomaly and novel results in non-local QED,''
JHEP \textbf{01}, 015 (2024)
\arXivold{arXiv}{2309.08417}{hep-ph}.


\bibitem{Lee:1969fy}
T.~D.~Lee and G.~C.~Wick,
``Negative Metric and the Unitarity of the S Matrix,''
Nucl. Phys. B \textbf{9}, 209-243 (1969)

\bibitem{Lee:1970iw}
T.~D.~Lee and G.~C.~Wick,
``Finite Theory of Quantum Electrodynamics,''
Phys. Rev. D \textbf{2}, 1033-1048 (1970)


\bibitem{Abu-Ajamieh:2024woy}
F.~Abu-Ajamieh, P.~Chattopadhyay and M.~Frasca,
``Phenomenological aspects of Lee-Wick QED,''
Nucl. Phys. B \textbf{1011}, 116799 (2025)
\arXivold{2406.16699}{hep-ph}.

\bibitem{Abu-Ajamieh:2024egb}
F.~Abu-Ajamieh, N.~Okada and S.~K.~Vempati,
``Aspects of non-local QED and the weak gravity conjecture,''
Eur. Phys. J. C \textbf{85}, no.5, 527 (2025)
\arXivold{2411.04877}{hep-ph}.



\bibitem{Dittrich:1998fy}
W.~Dittrich and H.~Gies,
``Light propagation in nontrivial QED vacua,''
Phys. Rev. D \textbf{58}, 025004 (1998)
\arXivold{hep-ph/9804375}{hep-ph}.

\bibitem{Kim:2021kif}
C.~M.~Kim and S.~P.~Kim,
``Magnetars as laboratories for strong field QED,''
AIP Conf. Proc. \textbf{2874}, no.1, 020013 (2024)
\arXivold{2112.02460}{astro-ph.HE}.



\bibitem{Denisov:2005si}
V.~I.~Denisov and S.~I.~Svertilov,
``Nonlinear electromagnetic and gravitational actions of neutron star fields on electromagnetic wave propagation,''
Phys. Rev. D \textbf{71}, 063002 (2005)

\bibitem{Denisov:2014oka}
V.~I.~Denisov, V.~A.~Sokolov and M.~I.~Vasili'ev,
``Nonlinear vacuum electrodynamics birefringence effect in a pulsar{\textquoteright}s strong magnetic field,''
Phys. Rev. D \textbf{90}, no.2, 023011 (2014)

\bibitem{Abishev:2014ceb}
M.~Abishev, Y.~Aimuratov, Y.~Aldabergenov, N.~Beissen, Z.~Bakytzhan and M.~Takibayeva,
``Some astrophysical effects of nonlinear vacuum electrodynamics in the magnetosphere of a pulsar,''
Astropart. Phys. \textbf{73}, 8-13 (2016)
\arXivold{1411.3127}{gr-qc}.

\bibitem{Denisov:2016pfu}
V.~I.~Denisov, E.~E.~Dolgaya and V.~A.~Sokolov,
``Nonperturbative QED vacuum birefringence,''
JHEP \textbf{05}, 105 (2017)
\arXivold{1612.09086}{hep-ph}.

\bibitem{Abishev:2018ahd}
M.~E.~Abishev, S.~Toktarbay, N.~A.~Beissen, F.~B.~Belissarova, M.~K.~Khassanov, A.~S.~Kudussov and A.~Z.~Abylayeva,
``Effects of non-linear electrodynamics of vacuum in the magnetic quadrupole field of a pulsar,''
Mon. Not. Roy. Astron. Soc. \textbf{481}, no.1, 36-43 (2018)


\bibitem{Muleri:2021wpd}
F.~Muleri, R.~Piazzolla, A.~Di Marco, S.~Fabiani, F.~La Monaca, C.~Lefevre, A.~Morbidini, J.~Rankin, P.~Soffitta and A.~Tobia, \textit{et al.}
``The IXPE instrument calibration equipment,''
Astropart. Phys. \textbf{136}, 102658 (2022)
\arXivold{2111.02066}{astro-ph.IM}.

\bibitem{eXTP:2018anb}
S.~N.~Zhang \textit{et al.} [eXTP],
``The enhanced X-ray Timing and Polarimetry mission{\textemdash}eXTP,''
Sci. China Phys. Mech. Astron. \textbf{62}, no.2, 29502 (2019)
\arXivold{1812.04020}{astro-ph.IM}.

\bibitem{HeylHernquist:1997}
J.~S.~Heyl and L.~Hernquist,
``Birefringence and dichroism of the QED vacuum,''
J. Phys. A \textbf{30}, 6485 (1997)
\arXivold{hep-ph/9705367}{hep-ph}.

\bibitem{ShavivHeylLithwick:1999}
N.~J.~Shaviv, J.~S.~Heyl and Y.~Lithwick,
``Magnetic lensing near ultramagnetized neutron stars,''
Mon. Not. Roy. Astron. Soc. \textbf{306}, 333 (1999)
\arXivold{astro-ph/9901376}{astro-ph}.

\bibitem{HeylShaviv:1999}
J.~S.~Heyl and N.~J.~Shaviv,
``Polarization evolution in strong magnetic fields,''
Mon. Not. Roy. Astron. Soc. \textbf{311}, 555 (2000)
\arXivold{astro-ph/9909339}{astro-ph}.

\bibitem{HeylShaviv:2002}
J.~S.~Heyl and N.~J.~Shaviv,
``QED and the high polarization of the thermal radiation from neutron stars,''
Phys. Rev. D \textbf{66}, 023002 (2002)
\arXivold{astro-ph/0203058}{astro-ph}.

\bibitem{LaiHo:2003}
D.~Lai and W.~C.~G.~Ho,
``Resonant conversion of photon modes due to vacuum polarization in a magnetized plasma,''
Astrophys. J. \textbf{588}, 962 (2003)
\arXivold{astro-ph/0108127}{astro-ph}.

\bibitem{HoLai:2003}
W.~C.~G.~Ho and D.~Lai,
``Atmospheres and spectra of strongly magnetized neutron stars II: The effect of vacuum polarization,''
Mon. Not. Roy. Astron. Soc. \textbf{338}, 233 (2003)
\arXivold{astro-ph/0201380}{astro-ph}.

\bibitem{vanAdelsbergLai:2006}
M.~van Adelsberg and D.~Lai,
``Atmosphere models of magnetized neutron stars: QED effects, radiation spectra, and polarization signals,''
Mon. Not. Roy. Astron. Soc. \textbf{373}, 1495 (2006)
\arXivold{astro-ph/0607168}{astro-ph}.

\bibitem{WangLai:2009}
C.~Wang and D.~Lai,
``Polarization evolution in a strongly magnetized vacuum: QED effect and polarized X-ray emission from magnetized neutron stars,''
Mon. Not. Roy. Astron. Soc. \textbf{398}, 515 (2009)
\arXivold{0903.2094}{astro-ph.HE}.

\bibitem{FernandezDavis:2011}
R.~Fern\'andez and S.~W.~Davis,
``The X-ray polarization signature of quiescent magnetars: effect of magnetospheric scattering and vacuum polarization,''
Astrophys. J. \textbf{730}, 131 (2011)
\arXivold{1101.0834}{astro-ph.HE}.

\bibitem{Taverna:2014}
R.~Taverna, F.~Muleri, R.~Turolla, P.~Soffitta, S.~Fabiani and L.~Nobili,
``Probing magnetar magnetosphere through X-ray polarization measurements,''
Mon. Not. Roy. Astron. Soc. \textbf{438}, 1686 (2014)
\arXivold{1311.7500}{astro-ph.HE}.

\bibitem{Kelly:2024}
R.~M.~E.~Kelly, S.~Zane, R.~Turolla and R.~Taverna,
``X-ray polarization in magnetar atmospheres: effects of mode conversion,''
Mon. Not. Roy. Astron. Soc. \textbf{528}, 3927 (2024)
\arXivold{2401.07618}{astro-ph.HE}.

\bibitem{Zane:2023}
S.~Zane \textit{et al.},
``A strong X-ray polarization signal from the magnetar 1RXS J170849.0-400910,''
Astrophys. J. Lett. \textbf{944}, L27 (2023)
\arXivold{2301.12919}{astro-ph.HE}.

\bibitem{TavernaTurolla:2024}
R.~Taverna and R.~Turolla,
``X-ray polarization from magnetar sources,''
Galaxies \textbf{12}, 6 (2024)
\arXivold{2402.05622}{astro-ph.HE}.

\bibitem{Stewart:2025}
R.~E.~Stewart \textit{et al.},
``Vacuum birefringence in the polarized X-ray emission of a radio magnetar,''
\arXivold{2509.19446}{astro-ph.HE}.

\bibitem{Taverna:2026}
R.~Taverna, R.~Turolla, L.~Marra, R.~M.~E.~Kelly, A.~Borghese, G.~L.~Israel, S.~Mereghetti, S.~Zane and M.~Rigoselli,
``The long quest for vacuum birefringence in magnetars: 1E 1547.0-5408 and the elusive smoking gun,''
\arXivold{2601.15452}{astro-ph.HE}.

\bibitem{eXTP:2025}
S.~N.~Zhang \textit{et al.},
``The enhanced X-ray Timing and Polarimetry mission---eXTP for launch in 2030,''
\arXivold{2506.08101}{astro-ph.HE}.

\bibitem{Abbassi:2026}
S.~Abbassi, F.~A.~Chishtie and S.~R.~Valluri,
``Finite-field QED corrections to vacuum birefringence and magnetar polarization transport,''
\arXivold{2607.06422}{hep-ph}.

\bibitem{Porto:2026}
G.~A.~Porto, J.~P.~Pereira, E.~Bittencourt and E.~Guzm\'an-Herrera,
``Nonlinear electrodynamics in magnetars: systematic effects on radius constraints and timing analysis,''
\arXivold{2605.17048}{gr-qc}.

\bibitem{GuoLai:2026}
T.~Guo and D.~Lai,
``Impact of resonant Compton scattering on magnetar X-ray polarization with QED vacuum resonance,''
\arXivold{2603.08119}{astro-ph.HE}.

\end{thebibliography}
\end{document}